\documentclass[twocolumn,showpacs,preprintnumbers,aps,prd,superscriptaddress,nofootinbib,10pt]{revtex4-1}
\usepackage{graphicx}
\usepackage{epsf}
\usepackage{bm}
\usepackage{amsmath}
\usepackage{amsfonts}
\usepackage{amssymb}
\usepackage{epstopdf}
\usepackage{natbib}
\usepackage{nameref}
\usepackage{color}
\usepackage{verbatim}
\usepackage{multirow}
\usepackage{bm}
\usepackage{dcolumn}
\usepackage{placeins}

\usepackage{subfigure}
\definecolor{darkblue}{rgb}{0.0, 0.0, 0.55}
\definecolor{darkred}{rgb}{0.55, 0.0, 0.0}
\usepackage{hyperref}
\hypersetup{
    colorlinks=true, 
    linkcolor=darkblue,
    citecolor=darkblue,
    urlcolor=darkblue}
        
\makeatletter\let\expandableinput\@@input\makeatother
\newcommand{\Omegam}{\Omega_{\mathrm{m}}}

\newcommand{\Omegab}{\Omega_{\mathrm{b}}}

\newcommand{\zd}{$z_\dagger$}

\begin{document}
\title{Statistical consistency of sign-switching vacuum energy with cosmological observations}
\author{Sehjal Khandelwal}
\email{sehjal.khandelwal@plaksha.edu.in}
\affiliation{Plaksha University, Mohali, Punjab-140306, India}

\author{Abraão J. S. Capistrano}
\email{capistrano@ufpr.br}
\affiliation{Universidade Federal do Paran\'{a}, Departamento de Engenharia e Exatas, Rua Pioneiro, 2153, Palotina, 85950-000, Paraná/PR, Brasil\\
Federal University of Latin American Integration (UNILA), Applied physics graduation program, Avenida Tarqu\'{i}nio Joslin dos Santos, 1000-Polo Universit\'{a}rio, Foz do Igua\c{c}u, 85867-670, Paran\'{a}/PR, Brasil}

\author{Suresh Kumar}
\email{suresh.kumar@plaksha.edu.in}
\affiliation{Plaksha University, Mohali, Punjab-140306, India}

\begin{abstract}
We assess dataset agreement and late-time predictive adequacy in $\Lambda$CDM and its sign-switching extension, $\Lambda_{\rm s}$CDM, using a suite of Gaussian and exact non-Gaussian consistency diagnostics. Both models are constrained with Cosmic Microwave Background (CMB) measurements from Planck, Atacama Cosmology Telescope (ACT), and South Pole Telescope (SPT), Baryon Acoustic Oscillation (BAO) data from Dark Energy Spectroscopic Instrument (DESI) Data Release 2 (DR2), and low-redshift Type Ia Supernova data from PantheonPlus+SH0ES. We find that commonly used Gaussian tension metrics can significantly overstate inconsistencies when broad, non-Gaussian posteriors are combined with tightly constrained datasets. In contrast, the exact non-Gaussian parameter shift indicates excellent consistency between the full CMB and BAO observations in both models. The $\Lambda_{\rm s}$CDM extension modestly improves geometric compatibility at intermediate redshifts, although reductions in parameter-level tension do not necessarily imply improved predictive consistency. These results highlight the importance of exact, non-Gaussian, and predictive diagnostics for robust assessments of cosmological model consistency.
\end{abstract}

\maketitle

\section{Introduction}
\label{Intro}
In the past twenty-five years, the amount and precision of cosmological data have increased significantly. The wide range of independent early and late universe probes constrains correlated cosmological parameters with unprecedented accuracy. Multiple independent probes \cite{2013PhR...530...87W}, observations of the Cosmic Microwave Background (CMB) anisotropies \cite{WMAP:2003elm, komatsu, Planck:2018vyg, madhavacheril}, Baryon Acoustic Oscillations (BAO) \cite{percival2007measuring, beutler,parkinson, alam2021completed, Alam2017}, Type-Ia Supernovae (SNe Ia) \cite{riess1998observational,perlmutter1999measurements} and weak gravitational lensing \cite{bartelmann2001weak,2015Liu,2018Shan,2018Mandelbaum, Heymans:2020gsg, DES:2021wwk} collectively establish the spatially flat six parameter $\Lambda$CDM model as a remarkable successful baseline description of the universe on large scales and confirmed the accelerated expansion of the universe. The standard $\Lambda$CDM model is identified with a cosmological constant, corresponding to a constant equation of state $w=-1$. Despite its remarkable success in describing a wide range of cosmological observations, the physical origin of dark energy remains as one of the most preferred puzzles in modern physics \cite{Weinberg:2008zzc, Ratra, Mortonson:2013zfa, Huterer:2017buf}. Again, when the predictions of $\Lambda$CDM are compared with late-time data, a number of enduring tensions have arisen as the accuracy of cosmological observations has gradually increased over time. One of the most urgent issues in modern cosmology is the difference between early and late universe estimates of the Hubble constant $H_0$, which is still statistically significant at the level of $\sim 5\!-\!7\sigma$. The measurement of the present-day expansion rate, i.e., the Hubble constant $H_0$, is one of the fundamental yet debated aspects of modern cosmology. Discrepancies in the measurements of $H_0$ from the early universe (e.g., CMB) and late universe (e.g., SNe Ia, BAO) have led to the Hubble tension \cite{Planck:2018vyg, Scolnic:2021amr, 2022ApJ...934L...7R, Freedman:2019jwv, Freedman:2024eph, H0DN:2025lyy, Verde:2019ivm, Abdalla:2022yfr, CosmoVerseNetwork:2025alb}. This has motivated a rich spectrum of theoretical extensions beyond $\Lambda$CDM model, refer to \cite{Kamionkowski:2022pkx, Najafi:2024qzm, CosmoVerseNetwork:2025alb, CANTATA:2021asi, Sotiriou:2008rp, DiValentino:2021izs} and the references therein.

Among these proposals are models in which the vacuum energy density deviates from strictly constant behavior at late times. Sign-switching or transitional vacuum energy scenarios, inspired by the graduated dark energy framework \cite{Akarsu:2019hmw} and related phenomenological considerations, enable the effective cosmological constant to grow from a negative value at earlier epochs to a positive value at later eras. Such models like the $\Lambda_{\rm{s}}$CDM model are one of the concrete realizations of the idea, in which the cosmological constant experiences an abrupt sign change at a characteristic transition redshift $z_\dagger$. This model has already been proposed and investigated in the literature  \cite{Akarsu:2021fol, Akarsu:2023mfb, Akarsu:2024eoo}, and it has been demonstrated to partially alleviate various late-time anomalies, notably tensions in $H_0$ and $S_8$, while remaining consistent with early-universe observables.

In the present work, we aim to provide a detailed and systematic assessment of the statistical consistency of the $\Lambda_{\rm{s}}$CDM framework when constrained using current cosmological datasets. As precision improves, it has become increasingly clear that apparent tensions between datasets can depend sensitively on the statistical tools used to quantify them. In particular, Gaussian approximations to posterior distributions can be misleading when posteriors are non-Gaussian, highly correlated, or when one dataset provides substantially weaker constraints than another \cite{Raveri_2021, Leizerovich_2024}. In such cases, simple comparisons based on parameter means and covariances may overstate or mischaracterize the degree of inconsistency between datasets. 

To address these issues, we adopt a comprehensive suite of dataset-consistency diagnostics discussed in section \ref{Stat}. In addition to standard parameter-shift estimators based on differences in posterior means (DM) and their updated variants (UDM), we consider goodness-of-fit-based statistics such as the difference in maximum a posteriori values (DMAP). We also employ the exact non-Gaussian parameter shift statistic, which is explicitly designed to remain valid in the presence of non-Gaussian posterior shapes. Complementing these parameter space diagnostics, we perform Posterior Predictive Consistency (PPC) tests \cite{Gelman:1996ppc, Feeney:2017sgx}. Posterior Predictive Distributions (PPD) have also been used to assess the internal consistency of cosmological probes. For example, the Dark Energy Survey (DES) Collaboration applied PPD-based tests to evaluate the mutual consistency of different probes in their joint cosmological analysis \cite{DES:2020lei}. This combination of methods allows us to disentangle genuine physical inconsistencies from artifacts arising from posterior geometry or dataset dominance. However, we restrict the PPC analysis to the low-redshift observables $(H_0,\mu)$ for clarity and interpretability. The main motivation is the late-time Hubble tension, for which a one-dimensional PPC on $H_0$ is the most direct predictive test. Furthermore, the low-redshift distance modulus $\mu(z=0.01)$ serves as a compact proxy for supernova distance information at the calibration end of the distance ladder, enabling a supplemental distance-based check without the difficulty of fully duplicating supernova datasets \cite{Gelman:1996ppc}. This targeted method allows us to determine whether models calibrated using CMB+LSS data can accommodate both local expansion-rate and nearby-distance limitations.

By applying this multi-pronged statistical analysis to both $\Lambda$CDM and its sign-switching extension $\Lambda_{\rm{s}}$CDM, we hope to clarify to what extent reported tension alleviations in the literature correspond to genuine improvements in predictive adequacy, rather than changes in parameter volume or posterior shape. This work establishes a reliable standard for evaluating late-time extensions of $\Lambda$CDM using present cosmological data.

The paper is organized as follows. In Sec. \ref{Models}, we outline the cosmological framework adopted in this work and describe the $\Lambda$CDM model together with its sign-switching extension, $\Lambda_{\rm s}$CDM. Sec. \ref{Datasets} summarizes the observational datasets used, including CMB measurements from Planck, Atacama Cosmology Telescope (ACT), and South Pole Telescope (SPT), BAO constraints from DESI DR2, and low-redshift supernova data from PantheonPlus+SH0ES. In Sec. \ref{method}, we detail the statistical methodology and consistency diagnostics employed to quantify the agreement of the dataset. Sec. \ref{Stat} presents the results of the tension metrics and posterior predictive consistency tests for both models. Finally, Sec. \ref{Result} discusses the implications of our findings, and Sec. \ref{conclusion} summarizes the main conclusions.

\section{Cosmological Framework and Models}
\label{Models}
We work within the standard relativistic framework of cosmology, assuming a homogeneous and isotropic universe described by the Friedmann-Lemaître-Robertson-Walker (FLRW) spacetime \cite{Weinberg:2008zzc, Peebles:1994xt, Ryden2017}. The cosmic expansion in the FLRW metric is governed by Einstein’s field equations \cite{Carroll:2004st}, which reduce to the Friedmann equation in a spatially flat universe (with $\rm c=1$) as
\begin{equation}
H^2=\frac{8\pi G}{3}\rho_{\rm{tot}},
\end{equation}
where $H$ is the Hubble expansion rate, $\rho_{\rm {tot}}$ denotes the total energy density of the cosmic fluid including baryonic matter, cold dark matter, radiation, and a cosmological constant or dark energy, also represented as $\Lambda$.

It is convenient to express the expansion rate in terms of redshift as
\begin{equation}
H^2(z)=H_0^2 \left[\Omega_{\rm{r0}}(1+z)^4+\Omega_{\rm{m0}}(1+z)^3+\Omega_{\rm{de0}}f(z)\right],
\end{equation}
where $\Omega_{\rm{r0}}$, $\Omega_{\rm{m0}}$, and $\Omega_{\rm{de0}}$ denote the present day density parameters of radiation, matter, and dark energy, respectively, and the function $f(z)$ encodes the redshift evolution of the dark-energy component \cite{Weinberg:2008zzc, AmendolaTsujikawa2010}. For a cosmological constant, corresponding to an equation of state, $w=-1$, one has $f(z)=1 \mathrm{(constant)}$.

The standard $\Lambda$CDM model is obtained by identifying dark energy with a constant $\Lambda$ and assuming spatial flatness. Despite its minimal assumptions, $\Lambda$CDM provides an excellent description of a wide range of cosmological observations as stated in section \ref{Intro}. The model is conventionally specified by six parameters describing the physical matter densities, the primordial perturbation spectrum, and the reionization history, and it assumes adiabatic initial conditions and a standard neutrino sector \cite{Planck:2018vyg}.

The other model considered in this work is the $\Lambda_{\rm{s}}$CDM model. This model is a minimal extension of the standard $\Lambda$CDM cosmology, introducing only a single additional parameter, \zd, at which the cosmological constant ($\Lambda$) undergoes an abrupt sign switch \cite{PhysRevD.109.103522}. All other standard cosmological components, including baryons, cold dark matter, pre-recombination physics, and the inflationary paradigm, remain unaltered. The sign-switching behavior is motivated by recent conjectures such as the graduated dark energy framework \cite{Akarsu:2019hmw}, which suggests a transition in the nature or behaviour of dark energy across cosmic time. Motivated by recent observational tensions, particularly in measurements of the Hubble constant \cite{Riess:2025chq, TDCOSMO:2025dmr}, the model introduces a phenomenological framework where the vacuum energy density is negative in the early universe and becomes positive at $z_\dagger\sim2$. This transition phenomenologically resembles an AdS-to-dS (Anti-de Sitter to de Sitter) phase shift, a behavior that has been discussed in various theoretical contexts, including effective descriptions motivated by high-energy frameworks \cite{Bousso:2000xa}. 
In its simplest realization, the cosmological constant is modeled as an abrupt sign-switching quantity \cite{Akarsu:2021fol, Akarsu:2022typ, Akarsu:2023mfb},
\begin{equation}
\Lambda \xrightarrow{} \Lambda_{\rm s} = \Lambda_{\rm s0}\,\mathrm{sgn}(z_\dagger - z),
\end{equation}
where $\Lambda_{\rm s0} > 0$ and $\mathrm{sgn}$ denotes the signum function, implying that the cosmological constant switches sign across the transition redshift $z_\dagger$.  
In the present work, we focus explicitly on this abrupt sign-switching realization of the $\Lambda_{\rm{s}}$CDM model \cite{Akarsu:2021fol, Akarsu:2022typ, Akarsu:2023mfb, Akarsu:2024eoo, Yadav:2024duq}. 
It is important to note, however, that the $\Lambda_{\rm{s}}$CDM scenario can be readily realized within the VCDM framework \cite{DeFelice:2020eju}, where the background dynamics are well defined over most redshift ranges, even though an exact description across all redshifts is not possible in the abrupt limit. In this broader context, smooth realizations of the sign transition can be constructed, as discussed for example in \cite{Akarsu:2024eoo}. See Refs. \cite{Alexandre:2023nmh, Anchordoqui:2023woo, Anchordoqui:2024dqc, Anchordoqui:2024gfa, Akarsu:2024nas, Souza:2024qwd, Akarsu:2025gwi, Akarsu:2024eoo, Soriano:2025gxd, Yadav:2024duq, Akarsu:2019hmw, Akarsu:2021fol, Akarsu:2023mfb, escamilla2025exploring} for explicit physical realizations of both abrupt and smooth versions of this framework.

Throughout this work, $\Lambda$CDM serves as the reference cosmology against which the statistical and physical performance of $\Lambda_{\rm{s}}$CDM is evaluated. This enables us to determine if the late-time sign-switching change enhances the internal consistency of cosmological datasets without sacrificing the standard $\Lambda$CDM framework's empirical success.

\section{Datasets}
\label{Datasets}
The datasets employed in this analysis are outlined below.

\textbf{Planck 2018:} We use the full Planck 2018 temperature and polarization measurements across both low and high multipoles\footnote{\url{https://github.com/benabed/clik}} \cite{Planck:2019cmb}. The high-$\ell$ likelihoods include the Planck TT, TE, EE spectra, covering the ranges $30 \leq \ell \leq 2508$ for TT and $30 \leq \ell \leq 1996$ for TE and EE. At large angular scales, we incorporate the low-$\ell$ temperature likelihood from Commander ($2 \leq \ell \leq 29$) and the low-$\ell$ polarization likelihood from SimAll ($2 \leq \ell \leq 29$). Additionally, we include the Planck lensing likelihood \cite{Planck:2018les} based on the SMICA reconstruction of the CMB lensing potential. These combined datasets provide a comprehensive set of temperature, polarization, and lensing constraints across the full multipole range probed by Planck \cite{Planck:2018vyg, Planck:2018les, Planck:2019cmb}.
The Planck dataset serves as the primary anchor for our cosmological constraints.

\textbf{ACT DR6:} We use the ACT DR6 ACT-lite likelihood\footnote{\url{https://github.com/ACTCollaboration/DR6-ACT-lite}}, which provides high-resolution TT, TE, and EE measurements over approximately 40\% of the sky and extends the small-scale sensitivity beyond Planck. In our analysis, we adopt the act\_dr6\_lenslike\footnote{\url{https://github.com/ACTCollaboration/act_dr6_lenslike}} configuration with multipole cuts of $\ell_{\max}=1000$ for TT and $\ell_{\max}=600$ for TE and EE. These data improve constraints in the damping-tail regime, complementing Planck's full-sky coverage with superior small-scale precision. The ACT-lite likelihood focuses on small angular scales, thereby extending the multipole range probed by Planck. The DR6 also includes Plik-lite likelihood cut at $\ell > 1000$ in TT, and $\ell > 600$ in TE and EE \cite{ACT:2023kun, AtacamaCosmologyTelescope:2025blo}.

\textbf{SPT-3G:} We use the SPT3G\_D1\_TnE\_lite\_candl likelihood\footnote{\url{https://github.com/SouthPoleTelescope/spt_candl_data/tree/main}} based on temperature and E-mode polarization power spectra from the SPT-3G experiment. The analysis employs the light compressed covariance without CMB lensing, along with internal priors and default data selection. This dataset complements Planck by improving sensitivity to small-scale CMB anisotropies, using publicly available TT, TE, and EE band-power data from ref\footnote{\url{https://pole.uchicago.edu/public/data/}} \cite{SPT:2023jql, SPT-3G:2021eoc, SPT-3G:2022hvq}.

This combined compilation of aforementioned CMB datasets is collectively referred to as the \textbf{CMB} in forthcoming text. Along with the full CMB dataset, two more datasets are used. 

\textbf{DESI DR2}: We use BAO measurements from Dark Energy Spectroscopic Instrument (DESI) Data Release 2 (DR2), including galaxy, quasar, and Lyman-$\alpha$ forest tracers across $0.295 \le z \le 2.330$, with distances reported as $D_{\rm M}/r_{\rm d}$, $D_{\rm H}/r_{\rm d}$, and $D_{\rm V}/r_{\rm d}$~\cite{DESI:2025dr2lya,DESI:2025dr2bao}.

\textbf{PPS}: We use the recently released Pantheon Plus compilation of Type Ia supernovae (SNe Ia). This release allows for the inclusion of low-redshift Cepheid calibration data from the SH0ES collaboration, which plays a key role in the absolute calibration of SNe Ia and is therefore directly relevant to studies of the Hubble tension. For clarity, we refer to this combined dataset as Pantheon Plus + SH0ES (\textbf{PPS}). The Pantheon Plus sample contains 1,701 SNe Ia spanning the redshift range $0.0012 <z<2.26$, as presented in\footnote{\url{https://github.com/PantheonPlusSH0ES/DataRelease}} \cite{Scolnic:2021amr}. 
\section{Methodology} 
\label{method}
We employ multiple consistency diagnostics because Gaussian tension measures can be biased by posterior non-Gaussianity and dataset dominance; the exact non-Gaussian shift and Posterior Predictive consistency (PPC) analysis\footnote{\url{https://github.com/sfeeney/hh0}} \cite{Feeney:2017sgx} provide robust cross-checks that quantify genuine parameter space inconsistency and predictive adequacy, respectively. We restrict the PPC analysis to $H_0$ and a low-redshift distance-modulus proxy $\mu(z=0.01)$ because these observables directly probe the late-time expansion history and encapsulate the dominant information relevant to the Hubble tension. Including higher-dimensional or early-universe observables in the PPC would largely reproduce the constraints already imposed by the CMB likelihoods and would therefore not provide an independent test of late-time predictive adequacy. Now, in order to perform the Markov Chain Monte Carlo (MCMC) analysis for both $\Lambda$CDM and $\Lambda_{\rm{s}}$CDM, we use the CLASS Boltzmann solver \cite{CLASS}. For the standard $\Lambda$CDM model, the publicly available CLASS implementation is employed\footnote{\url{https://github.com/lesgourg/class_public}}. For the $\Lambda_{\rm{s}}$CDM scenario, we modify the CLASS background module to implement a non-clustering, sign-switching cosmological constant. In this framework, in addition to the six baseline $\Lambda$CDM parameters, we introduce a new parameter \zd, which denotes the redshift at which the effective cosmological constant changes sign. Specifically, the cosmological constant is negative at early times and positive at late times, corresponding to an AdS-like and a dS-like effective background evolution, respectively. The choice of the cosmological constant in the $\Lambda_{\rm{s}}$CDM model modifies only the homogeneous background expansion history and assumes that dark energy does not cluster. The step-function behavior serves as a simplifying approximation to a rapid but continuous sign transition. With this modification, the Friedmann equation takes the form
\begin{equation}
\frac{H^2(z)}{H_0^2}
= \Omega_{\rm {r0}}(1+z)^4
+ \Omega_{\rm{m0}}(1+z)^3
+ \Omega_{\rm {\Lambda_{\rm{s}} 0}}\,\mathrm{sgn}(z_\dagger - z),
\end{equation} where $\Omega_{\rm{r0}}$, $\Omega_{\rm{m0}}$, and $\Omega_{\rm {\Lambda_{\rm{s}} 0}}$ denote the present-day density parameters for radiation, matter, and the sign-switching cosmological constant, respectively. All present-day density parameters retain their standard definitions, with the critical density given by $\rho_{\rm{c0}} = 3H_0^2/(8\pi G)$ and
\begin{equation}
\Omega_{\rm{r0}} = \frac{\rho_{\rm{r0}}}{\rho_{\rm{c0}}}, \qquad \Omega_{\rm{m0}} = \frac{\rho_{\rm{m0}}}{\rho_{\rm{c0}}}, \qquad
\Omega_{\rm{\Lambda_{\rm{s}} 0}} = \frac{\rho_{\rm{\Lambda_{\rm{s}} 0}}}{\rho_{\rm{c0}}}.
\end{equation}
On sub-horizon scales, since dark energy perturbations are neglected, the growth of matter density perturbations remains governed by the standard equation 
\begin{equation}
\ddot{\delta} + 2H\dot{\delta} - 4\pi G \rho_{\rm{m}} \delta = 0,
\end{equation}
ensuring that structure formation is affected only indirectly through the modified background expansion \cite{Ma:1995ey, Bouhmadi-Lopez:2025spo, 1992PhR...215..203M}.

The MCMC chains generated using this modified CLASS implementation are subsequently analyzed using the \texttt{Tensiometer}\footnote{\url{https://github.com/mraveri/tensiometer}} framework to quantify dataset consistency and statistical tensions \cite{Leizerovich:2023qqt}. The cosmological constraints are obtained by running MCMC simulations using the CLASS Boltzmann solver interfaced with the MontePython framework\footnote{\url{https://github.com/brinckmann/montepython_public}}. We explore the full parameter space of both the $\Lambda$CDM and $\Lambda_{\rm{s}}$CDM models using the Metropolis-Hastings sampling algorithm \cite{MH2015arXiv150401896R}. The convergence of the chains is verified using the Gelman-Rubin criterion, $R-1<0.01$ \cite{GelmanRubin1992, Gellmaan_R}. We perform independent MCMC analyses for the three datasets employed in this work: CMB, DESI DR2, and PantheonPlus+SH0ES (PPS). The resulting posterior samples are used to compute a range of statistical tension metrics, including the parameter difference (DM), update difference in mean (UDM), maximum a posteriori difference ($D_{\mathrm{MAP}}$), the exact non-Gaussian parameter shift, and the PPC consistency test. The tension metrics are evaluated using the \texttt{Tensiometer} package, which operates directly on the MCMC chains. The analysis is performed both on the full shared parameter set to assess the robustness of Gaussian approximations. We additionally evaluate the effective number of constrained parameters and goodness of fit statistics \cite{2019arXiv191201134S} to diagnose posterior non-Gaussianities and dataset dominance effects, ensuring that apparent tensions are not driven by mismatched posterior shapes. We do not perform a separate prior-only MCMC run. All tension metrics are computed directly from the posterior chains obtained from the data analyses, assuming standard, weakly informative priors on the sampled parameters. Since our analysis focuses on posterior-level dataset consistency rather than prior–posterior volume effects, a dedicated prior-only chain is not required. We adopt uniform (flat) priors on the full set of cosmological parameters: the physical baryon density $\omega_{\rm b}\in[0.018,0.024]$, the physical cold dark matter density $\omega_{\rm c}\in[0.10,0.14]$, the angular size of the sound horizon at recombination parameterized by $100\theta_{\rm s}\in[1.03,1.05]$, the logarithmic amplitude of the primordial scalar power spectrum $\ln(10^{10}A_{\rm s})\in[3.0,3.18]$, the scalar spectral index $n_{\rm s}\in[0.9,1.1]$, the optical depth to reionization $\tau_{\rm reio}\in[0.04,0.125]$, and the additional redshift parameter $z_\dagger\in[1,3]$. All prior ranges are chosen to be sufficiently large to ensure that the resulting parameter constraints are dominated by the data rather than by prior-volume effects.

\section{Statistical Analysis and Results for Various Tension Metrics}
\label{Stat}
In this section, we describe the different tension metrics and present the results of our analysis. The datasets compared are CMB, DESI DR2, and PPS for the two models, $\Lambda$CDM and $\Lambda_{\rm s}$CDM. Before presenting the final results and discussion, we first summarize the general behavior and results of all the metrics.

\subsection{Rule Of Thumb}
This section describes the different tension metrics as proposed by \cite{Raveri_2019, Raveri_2021} on the basis of Bayesian inference. The metrics are discussed based on the posterior distributions $P(\theta|D)$ for model M, which is implemented in the tensiometer\footnote{\url{https://github.com/mraveri/tensiometer}}. All the findings are reported in terms of the effective number of standard deviation $N_\sigma$ for a two-sided Gaussian probability, described by \cite{Raveri_2019} as
\begin{equation}
    P = \operatorname{erf}\left(\frac{N_\sigma}{\sqrt{2}}\right) .
    \label{sigma}
\end{equation}
where $P$ is the likelihood of agreement or disagreement across datasets. Furthermore, $ P = 1-\rm PTE$ \cite{Leizerovich_2024} where PTE is the probability to exceed a certain observed value of the estimator $Q$ noted as $Q*$ for all different metrics, respectively.

In current cosmology, measurements from distinct data sets (for example, CMB vs. large-scale structure PPS) might result in somewhat different values for the same cosmological parameter, such as the Hubble constant. Simple rules of thumb are often used to evaluate these ``tensions'' before more robust statistical tests are performed \cite{DES:2020hen,2019NatAs...3..891V,  Freedman:2021ahq}. The thumb rule given in \cite{Lemos_2021, PhysRevD.100.043504} states that in order to compare two independent measurements of a parameter \(\theta \), just calculate the difference in their mean values, normalized by the quadratic sum of their errors. Therefore, the tension metric is given by:
\begin{equation}
T_{1}=\frac{|\theta _{1}-\theta _{2}|}{\sqrt{\sigma _{1}^{2}+\sigma _{2}^{2}}}
\label{Rule_of_thumb}
\end{equation}
A value of \(T_{1}\) more than \(3\sigma \) (standard deviations) generally causes alarm and needs additional study into the data, or underlying physics.
Another way to determine if two distinct datasets can be defined by the same set of cosmological parameters using their posterior distributions is by determining various tension metrics as discussed in \cite{Raveri_2019, Leizerovich_2024}. These essentially compare the odds of two distinct statements: 1) The two datasets $D_1$ and $D_2$ are consistent, i.e., they are defined using the same set of parameters. 2) The two datasets $D_1$ and $D_2$ are not consistent for the given model, respectively.

\subsection{Posterior Means of Two Datasets in Standard Form}
The difference between the posterior means of the two datasets $D_1(\theta)$ and $D_2(\theta)$ for the inferred parameters is determined using the quadratic form defined as \cite{Leizerovich_2024, Raveri_2019}
\begin{equation}
Q_{\text{DM}} = (\hat{\theta}_{\rm D_1} - \hat{\theta}_{\rm D_2})^{T} C^{-1} (\hat{\theta}_{\rm D_1} - \hat{\theta}_{\rm D_2})
\label{DM}
\end{equation}
\noindent
where $C = \hat{C}_{\rm D_1} + \hat{C}_{\rm D_2}$ is the covariance matrix of $(\hat{\theta}_{\rm D_1} - \hat{\theta}_{\rm D_2})$. $\hat{\theta_i}$ and $\hat{C_i}$ are the mean and covariance matrix on the parameter space for datasets $D_1$ and $D_2$. 
With the Gaussian choice of posterior distribution, the quadratic form of $Q_{\rm{DM}}$ follows a $\chi^2$ distribution with $\nu = \text{rank}[\hat{C}_{\rm D_1}+\hat{C}_{\rm D_2}]$ degrees of freedom (dof) \cite{Raveri_2019} respectively.  

\begin{table}[hbp]
\centering
\caption{Parameter Shift Consistency Results for $\Lambda$CDM Model Using CMB+DESI DR2 Dataset Combination in full shared parameter set. In the table below $\rm UDM_{1}$ refer to $\rm{UDM}_{(\rm{CMB\xrightarrow{}DESI DR2})}$ and $\rm{UDM}_{2} $ refer to $\rm{UDM}_{(\rm{DESI DR2\xrightarrow{}CMB})}$, respectively. $P$ is truncated to two decimals in the Tables; PTE and $N_{\sigma}$ are the authoritative values.}
\label{tab:lcdm_base_CMB}
\begin{tabular}{lccccc}
\hline\hline
\textbf{Metric} & $\nu$ & $Q^*$ & $P$ & PTE & $N_\sigma$ \\
\hline
DM     & 4 & 11.2 & 0.98  & $0.022$  & 2.25 \\
$\rm{UDM}_{1}$ & 2 & 4.5  & 0.9  & 0.1  & 1.6 \\
$\rm{UDM}_{2}$ & 3 & 21.9 & 0.99 & $6.7\times10^{-5}$ & 3.98 \\
DMAP   & 4 & 6.4  & 0.83  & 0.17  & 1.4 \\
Exact Param. Shift & -- & -- & 0.97 & 0.026 & 2.2 \\
\hline
\multicolumn{6}{l}{%
\begin{tabular}{@{}l@{\hspace{1em}}l@{}}
\textbf{Rule of Thumb:} & $N_{\Omegab}=0.0914\quad N_{\Omegam}=2.00$ \\
                        & $N_{H_0}=2.495\quad N_{r_{d}}=0.5931$
\end{tabular}}\\
\hline\hline
\end{tabular}
\end{table}

\begin{table}[tbp]
\centering
\caption{Parameter Shift Consistency Results for $\Lambda$CDM Model Using DESI DR2+CMB+PPS Dataset Combination in full shared parameter set. In the table below $\rm UDM_{1}$ refer to $\rm{UDM}_{(\rm{CMB+DESI DR2\xrightarrow{}PPS})}$ and $\rm{UDM}_{2} $ refer to $\rm{UDM}_{(\rm{PPS\xrightarrow{}CMB+DESI DR2})}$ respectively. $P$ is truncated to two decimals in the Tables; PTE and $N_{\sigma}$ are the authoritative values.}
\label{tab:lcdm_base_CMB_pp}
\begin{tabular}{lccccc}
\hline\hline
\textbf{Metric} & $\nu$ & $Q^*$ & $P$ & PTE & $N_\sigma$ \\
\hline
DM     & 4 & 39.8 & 1 & $4.6\times10^{-8}$ & 5.5 \\
$\rm{UDM}_{1}$ & 2 & 1.2  & 0.45 & 0.6 & 0.58 \\
$\rm{UDM}_{2}$ & 2 & 11.35 & 0.99 & $3.4\times10^{-3}$ & 2.92 \\
DMAP   & 4 & 20 & 0.99 & $0.0005$ & 3.5 \\
Exact Param. Shift & -- & -- & $< 2\times 10^{ -7}$ & 0.999 & $\geq 5.1$ \\
\hline
\multicolumn{6}{l}{%
\begin{tabular}{@{}l@{\hspace{1em}}l@{}}
\textbf{Rule of Thumb:} & $N_{\Omegab}=0.2678\quad N_{\Omegam}=1.33$ \\
                        & $N_{H_0}=5.6784\quad N_{r_{d}}=3.7417$
\end{tabular}}\\
\hline\hline
\end{tabular}
\end{table}

\begin{table}[tbp]
\centering
\caption{Parameter Shift Consistency Results for $\Lambda_{\rm{s}}$CDM Model Using CMB+DESI DR2 Dataset Combination in full shared parameter set). In the table below $\rm UDM_{1}$ refer to $\rm{UDM}_{(\rm{CMB\xrightarrow{}DESI DR2})}$ and $\rm{UDM}_{2} $ refer to $\rm{UDM}_{(\rm{DESI DR2\xrightarrow{}CMB})}$ respectively. $P$ is truncated to two decimals in the Tables; PTE and $N_{\sigma}$ are the authoritative values.}
\label{tab:lcdm_full_base_CMB}
\begin{tabular}{lccccc}
\hline\hline
\textbf{Metric} & $\nu$ & $Q^*$ & $P$ & PTE & $N_\sigma$ \\
\hline
DM     & 6 & 1.6 & 0.05 & 0.95 & 0.061 \\
$\rm{UDM}_{1}$ & 3 & 0.95 & 0.185 & 0.81 & 0.23 \\
$\rm{UDM}_{2}$ & 4 & 0.33 & 0.0125 & 0.99 & 0.016 \\
DMAP   & 6 & 2.98 & 0.19 & 0.81 & 0.24 \\
Exact Param. Shift & -- & -- & 0.66 & 0.34 & 0.95 \\
\hline
\multicolumn{6}{l}{%
\begin{tabular}{@{}l@{\hspace{1em}}l@{\hspace{1em}}l@{}}
\textbf{Rule of Thumb:} & $N_{\Omegab}=0.0328\quad N_{\Omegam}=0.50$ \\
                        & $N_{H_0}=0.70\quad N_{r_{d}}=0.133$ \\
                        & $N_{P1}=0.75$\quad $N_{\Omega_{\rm{cdm}}}=0.098$
\end{tabular}}\\
\hline\hline
\end{tabular}
\end{table}

\begin{table}[tbp]
\centering
\caption{Parameter Shift Consistency Results for $\Lambda_{\rm{s}}$CDM Model Using DESI DR2+CMB+PPS Dataset Combination in full shared parameter set). In the table below $\rm UDM_{1}$ refer to $\rm{UDM}_{(\rm{CMB+DESI DR2\xrightarrow{}PPS})}$ and $\rm{UDM}_{2} $ refer to $\rm{UDM}_{(\rm{PPS\xrightarrow{}CMB+DESI DR2})}$ respectively. $P$ is truncated to two decimals in the Tables; PTE and $N_{\sigma}$ are the authoritative values.}
\label{tab:lcdm_full_base_CMB_pp}
\begin{tabular}{lccccc}
\hline\hline
\textbf{Metric} & $\nu$ & $Q^*$ & $P$ & PTE & $N_\sigma$ \\
\hline
DM     & 6 & 77.4 & 1 & $1.24\times10^{-14}$ & 7.7 \\
$\rm{UDM}_{1}$ & 3 & 0.6  & 0.103 & 0.9 & 0.13 \\
$\rm{UDM}_{2}$ & 3 & 67.59 & 1 & $1.40\times10^{-14}$ & 7.6 \\
DMAP   & 6 & 15.7 & 0.98 & 0.015 & 2.42 \\
Exact Param. Shift & -- & -- & $< 2\times 10^{ -7}$   & 0.999 & $\geq 5.1$ \\
\hline
\multicolumn{6}{l}{%
\begin{tabular}{@{}l@{\hspace{1em}}l@{\hspace{1em}}l@{}}
\textbf{Rule of Thumb:} & $N_{\Omegab}=0.0378\quad N_{\Omegam}=1.1430$ \\
                        & $N_{H_0}=3.9508\quad N_{r_{d}}=2.8155$ \\
                        & $N_{P1}=0.9159$\quad $N_{\Omega_{\rm{cdm}}}=2.7893$
\end{tabular}}\\
\hline\hline
\end{tabular}
\end{table}

\begin{figure*}[t]
\centering

\subfigure[]{
    \includegraphics[width=0.48\textwidth]{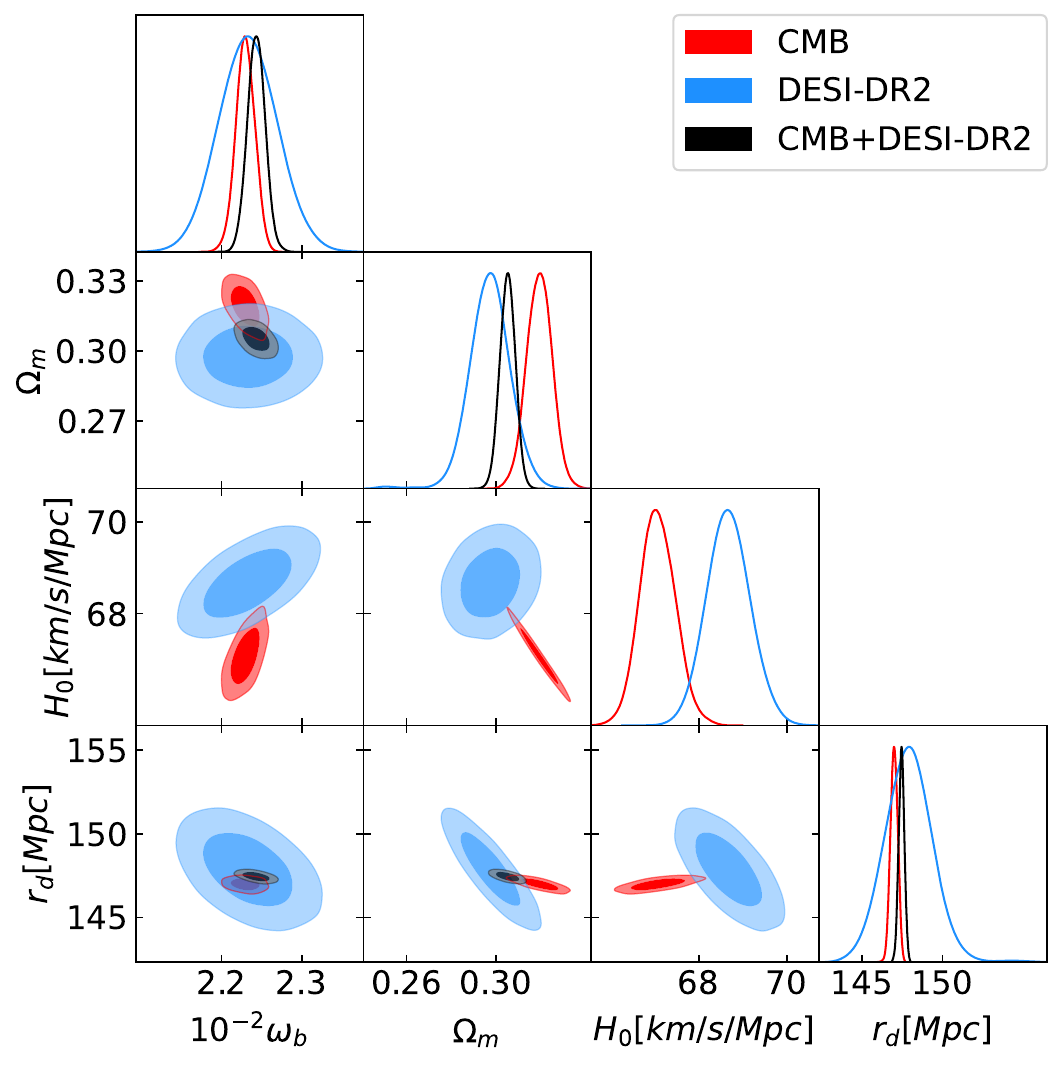}
    \label{fig1}
}
\hfill
\subfigure[]{
    \includegraphics[width=0.48\textwidth]{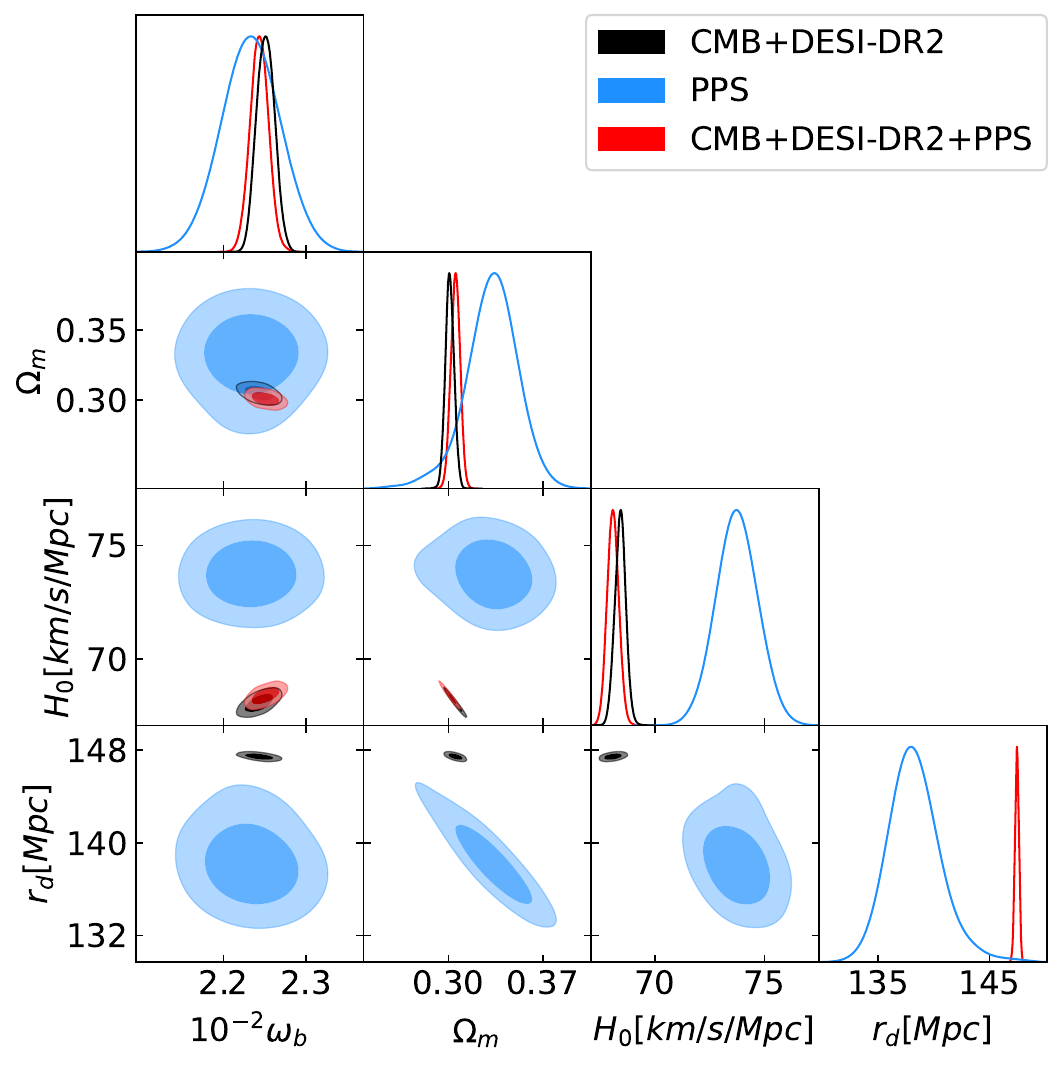}
    \label{fig2}
}

\caption{Triangular plots for the $\Lambda$CDM model. 
(Left) Comparison between CMB (red), DESI DR2 (blue), 
and their combination (black). (Right) Comparison between the combined CMB+DESI DR2 dataset (black), PPS (blue), and the full combination (red). Diagonal panels show 1D posteriors; off-diagonal panels show 68\% and 95\% contours.}
\label{fig:lcdm_panels}
\end{figure*}

According to Table~\ref{tab:lcdm_base_CMB}, the ${\Lambda}$CDM shows a weak to moderate tension between DESI DR2 and CMB, consistent with the moderate mean differences seen in Fig. \ref{fig1}. From Table~\ref{tab:lcdm_base_CMB_pp}, the strongest tension arises between the joint CMB+DESI~DR2 dataset and PPS, which means the two datasets assume two different parameter values in this model.

\begin{figure*}[t]
\centering

\subfigure[]{
    \includegraphics[width=0.48\textwidth]{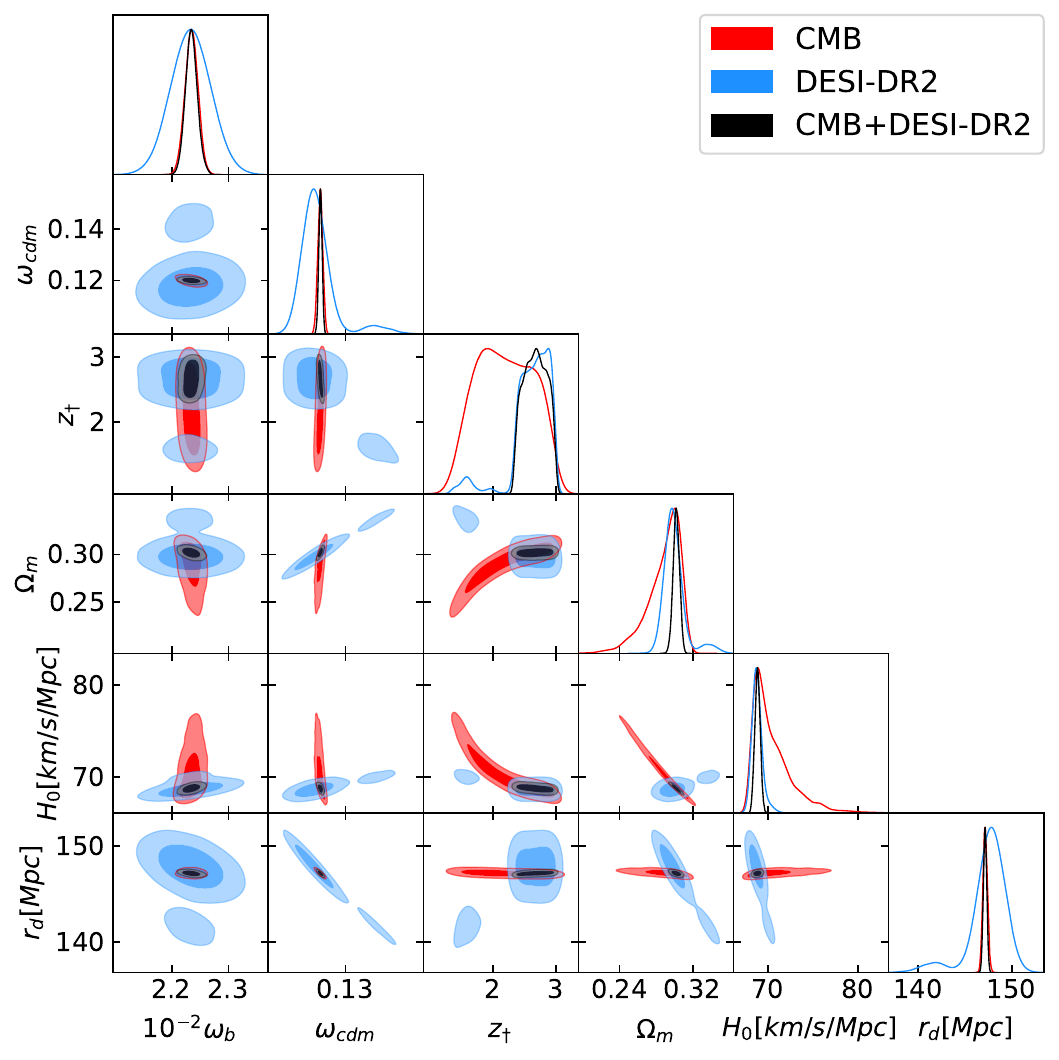}
    \label{fig3}
}
\hfill
\subfigure[]{
    \includegraphics[width=0.48\textwidth]{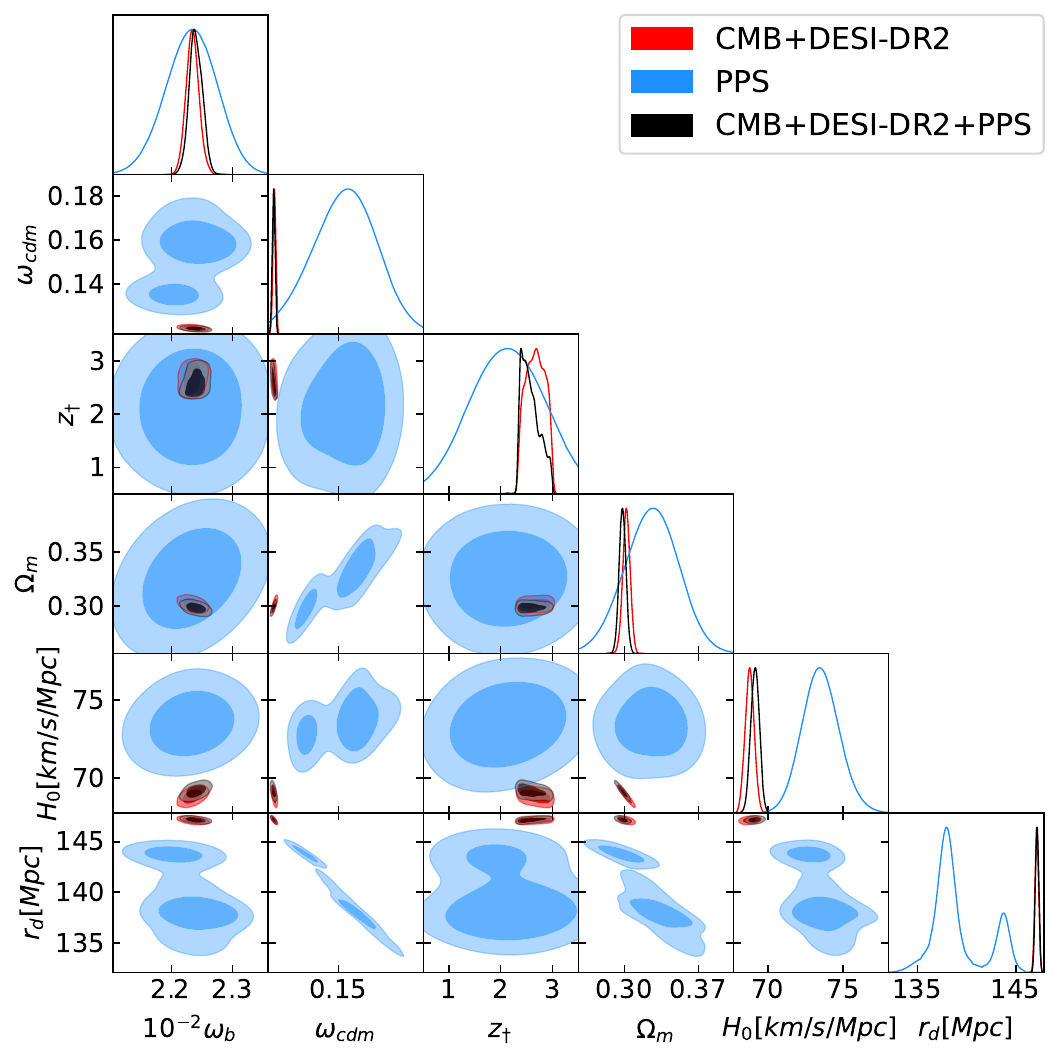}
    \label{fig4}
}

\caption[]{Triangular plots for the $\Lambda_{\rm s}$CDM model.
(Left) Comparison between CMB (red), DESI DR2 (blue), and their combination (black).
(Right) Comparison between CMB+DESI DR2 (red), PPS (blue), and the full combination (black).
Diagonal panels show 1D posteriors; off-diagonal panels show 68\% and 95\% credible regions.}
\label{fig:lscdm_panels}
\end{figure*}

In the ${\Lambda_{\rm s}}$CDM model, DESI DR2 and CMB represent the lowest tension among all analyses (Table~\ref{tab:lcdm_full_base_CMB} and Fig. \ref{fig3}). The comparison between CMB+DESI DR2 and PPS yields an extremely large tension of $7.7\sigma$, driven primarily by the Gaussian approximation assumed in the DM metric. We point out that although the Gelman–Rubin criterion 
$R-1 \ll 0.01$ is satisfied, we reinforce that $R$ is a necessary but not sufficient condition for full exploration, particularly for extended cosmological models where posteriors can be highly non-Gaussian and exhibit long degeneracy directions. In such cases, multiple Metropolis–Hastings chains may show small between-chain variance (hence $R\sim 1$) while still mixing slowly along weakly constrained directions, leading to residual Monte Carlo noise visible in marginal/triangle plots. This behavior is consistent with the observation that DESI DR2 in ${\Lambda_{\rm s}}$CDM is comparatively less convergent and that DR2 alone does not fully converge when used independently. To ensure robustness of our results, we additionally verified stability of the constraints with adequate effective sample sizes (ESS) for the parameters. The bulk ESS is $\sim 6\times 10^2- 8\times 10^2$ and tail ESS is $\sim 1.1\times 10^3-1.4\times 10^3$, indicating that medians and $68\%$ C.I. are stable to Monte Carlo error. The largest rank-split $R$ we find is $\sim 1.014$ for $\omega_{cdm}$, consistent with mild residual slow mixing but not affecting our results.

It is important to note that DM accounts for parameter correlations, while the rule of thumb does not, leading to different tension estimates \cite{Raveri_2019}.

\subsection{Posterior Means of Two Datasets in Updated Form}
This quantity is referred to as the updated difference in mean since it quantifies the differences in parameters of one dataset $D_1$ and the joint dataset $D_{12}$ referred to as the UDM metric \cite{Leizerovich_2024}. The updated difference in the mean quadratic form is:
\begin{equation}
Q_{\text{UDM}} = (\hat{\theta}_{\rm D_1} - \hat{\theta}_{\rm D_{12}})^{T} C^{-1} (\hat{\theta}_{\rm D_1} - \hat{\theta}_{\rm D_{12}})
\label{UDM}
\end{equation}
\noindent
where $C = \hat{C}_{\rm D_1} - \hat{C}_{\rm D_{12}}$ is the covariance matrix of $(\hat{\theta}_{\rm D_{1}} - \hat{\theta}_{\rm D_{12}})$. $\hat{\theta}_{\rm D_{12}}$ and $\hat{C}_{\rm D_{12}}$ are the mean and covariance matrix obtained from the statistical analysis with the joint dataset $D_{{12}}$.\\ 
All these quadratic forms are central and can be degenerate, depending on the covariance rank. For Gaussian posteriors of both the datasets $D_1$ and $D_{12}$, $Q_{\rm{UDM}}$ follows a chi squared distribution with rank $\nu = \text{rank}[\hat{C}_{\rm D_1}-\hat{C}_{\rm D_{12}}]$ dof. The UDM statistic is directional in principle as it measures how much the updated dataset shifts the posterior mean of the base dataset, and this depends on which dataset plays each role \cite{Raveri_2020}. It has applications in non-Gaussian metrics.

In the $ \Lambda$CDM model, updating CMB with DESI DR2 yields a weak tension of $1.6\sigma$ indicates a smaller shift, whereas when DESI DR2 is updated with CMB, the obtained tension increases to $3.98\sigma$ implies that the CMB dataset is more consistent internally. According to Eq. \ref{UDM}, the tension is higher if the mean shift is higher or $|\hat C_{\rm D_{1}} - \hat C_{\rm D_{12}}|$ is small. Here, the covariance matrix is quite small, and the posterior mean is non-negligible, making the $Q^*$ estimator (or $Q_{\rm UDM}$ estimator) and corresponding $N_\sigma$ higher. Furthermore, $\rm {UDM}_2$ in Table \ref{tab:lcdm_base_CMB} has 3 effective degrees of freedom. Since DESI DR2 alone is relatively weak, CMB data add a lot of information and significantly tighten three independent combinations of parameters. A similar behavior appears with the PPS dataset.  Updating CMB+DESI DR2 with PPS gives a negligible tension of $0.58\sigma$, while updating PPS with the combined data increases the tension to $2.92\sigma$. This asymmetric tension arises because PPS has an extremely large posterior; it covers almost the entire region of the combined dataset's posterior, and the update dataset (CMB+DESI DR2) forces PPS into a much narrower parameter region.

In the $ \Lambda_{\rm{s}}$CDM model, we first consider the case of CMB+DESI DR2. Updating CMB with DESI DR2 yields a tension of $0.23\sigma$, while updating DESI DR2 with CMB gives $0.016\sigma$, thus it substantially decreases. Both tensions are negligible, consistent with Fig. \ref{fig3}, where the contours overlap closely, with DESI DR2 appearing less convergent. Next, we analyze the CMB+DESI DR2 and PPS comparison. Updating CMB+DESI DR2 with PPS gives a tension of $0.13\sigma$, whereas updating PPS with the combined dataset results in an extremely strong tension of $7.6\sigma$, which means the posterior mean is dragged a long distance, far beyond what PPS would ``expect'' statistically, inducing tension in between the datasets. In this case, the difference between the means of the two distributions becomes significant.

\subsection{Goodness-of-fit type test}
Using this metric, we evaluate the difference of log-likelihoods at the Maximum A Posteriori (MAP) point for two datasets considered jointly and separately in statistical analysis using the estimator defined in \cite{Leizerovich_2024, Raveri_2019}
\begin{equation}
Q_{\mathrm{DMAP}} = -2 \ln \mathcal{L}_{\mathrm{12}}(\theta_p^{12}) 
+ 2 \ln \mathcal{L}_1(\theta_p^{1}) 
+ 2 \ln \mathcal{L}_2(\theta_p^{2}) 
\label{DMAP}
\end{equation}
$\theta_{p}^{i}$ denotes the MAP parameters for dataset $i = (1,2,12)$. This statistic quantifies the loss in goodness of fit when two independent datasets are combined \cite{Raveri_2019}. The normalization factors of the likelihoods cancel for the independent datasets, ensuring that $Q_{\mathrm{DMAP}}$ depends only on the shape of the likelihood surfaces and not on their absolute normalization. Therefore, the resulting value of $Q_{\mathrm{DMAP}}$ measures how much statistical tension exists between the two datasets. Under the assumption of Gaussian priors, the distribution of $Q_{\mathrm{DMAP}}$ can be approximated as $Q_{\mathrm{DMAP}} \sim \chi^2( N_{\mathrm{eff}}^{(1)} + N_{\mathrm{eff}}^{(2)} - N_{\mathrm{eff}}^{(12)})$, where the sum of $N_{\mathrm {eff}}$ is the number of dof \cite{Raveri_2019}, and the effective number of constrained parameters for each dataset is defined as
\begin{equation}
N_{\mathrm{eff}}^{(i)} = N - \mathrm{tr} \left[ C_{\Pi,i}^{-1} \, C_{p,i} \right] 
\end{equation}
where $N$ is the number of sampled parameters. $C_{\Pi, i}$ and $C_{p,i}$ are the covariance matrices of the posterior and prior distributions for the $i$-datasets, respectively \cite{Spiegelhalter:2002yvw, Raveri_2019, Raveri_2021, Raveri_2020, Lemos_2021, Leizerovich:2023qqt, Leizerovich_2024}. Therefore, $Q_{\mathrm{DMAP}}$ serves as a useful tool to measure how well the data fit the theoretical predictions, though it does not provide a direct estimate of the inconsistency between individual parameters. DMAP cannot be compared to DM because, unlike DM, DMAP does not take the covariance matrix into account.

In the $ \Lambda$CDM model, for the joint CMB+DESI DR2 dataset, DMAP gives $1.4\sigma$ tension, while DESI DR2+CMB+PPS gives $3.4\sigma$ tension, indicating that the broad posterior of PPS pulls the best–fit value away from the CMB+DESI DR2 solution. The results are in agreement with Fig. \ref{fig1} and Fig. \ref{fig2}. Since flat uninformative priors are used, the effective degrees of freedom reported in Tables \ref{tab:lcdm_base_CMB}-\ref{tab:lcdm_full_base_CMB_pp} correspond to the numerical rank of the parameter-difference covariance matrix and therefore may be smaller than the total number of sampled parameters. We also point out that with bounded priors, $N_{\mathrm{eff}}$ depends on how much the posterior shrinks relative to those bounds; thus, it will not be equal to the dimension.

In the $\Lambda_{\rm{s}}$CDM model, in Fig. \ref{fig3}, we can see that the maximum of the joint posterior is located in the $1\sigma$ contour of the corresponding posteriors obtained considering only one dataset. Similarly, in Fig. \ref{fig4}, we can see that the maximum of the joint posterior is located in the $1\sigma$ contour of CMB+DESI DR2 posterior, except for the $H_0$ and $r_d$ value which, lies in the $2\sigma$ contour. Thus, this gives rise to a mild tension as quantified in Table \ref{tab:lcdm_full_base_CMB_pp}. Both the observations are in agreement with the results quantified in Tables \ref{tab:lcdm_full_base_CMB} and \ref{tab:lcdm_full_base_CMB_pp}. Note that the tensions obtained in all the above cases, for both models, remain lower than the $H_0$ tension estimated by the rule of thumb. This indicates that the overall tension decreases when the full shared parameter space is taken into account.

It is important to point out that in our analysis, we adopt weakly informative uniform priors over ranges that are substantially broader than the final posterior support. Because the posterior distributions are strongly data-dominated within these bounds, prior-volume effects are subdominant for the primary tension estimators. That said, we emphasize that $N_{\mathrm{eff}}$ should not be interpreted as the literal dimensionality of the parameter space, but as a measure of posterior contraction relative to the adopted prior domain. Moreover, due to the fact that the priors are chosen to be sufficiently broad and encompass regions far beyond the posterior bulk, the resulting  $N_{\mathrm{eff}}$ values primarily reflect data-driven constraints. Thus, while prior-only chains could in principle refine the interpretation of $N_{\mathrm{eff}}$, their omission does not affect the parameter-shift or posterior predictive conclusions presented here. The core tension diagnostics are computed directly from posterior samples and remain robust under variations of prior width within reasonable bounds.

\subsection{Exact Parameter Shift in Non-Gaussian Metrics}
To robustly quantify dataset consistency, we employ the non-Gaussian exact parameter shift statistic introduced in \cite{Raveri_2019} and further used in \cite{Raveri_2021}. This method evaluates the parameter difference probability density $P(\Delta\theta)$, where $\Delta\theta=\theta_{1}-\theta_{2}$ is the difference between the means of the posterior parameters that correspond to datasets $1/2$. The general expression for two uncorrelated datasets is given by \cite{Leizerovich_2024}
\begin{equation}
    P(\Delta\theta) = \int P_{A}(\theta)P_{B}(\theta-\Delta\theta)\,d \theta\, ,
    \label{eq: Delta_theta}
\end{equation}
The probability to identify tension or statistical significance of the shift is calculated corresponding to the probability that the parameter shift exceeds zero $\Delta\theta = 0$. Therefore, $\Delta = \int_{P(\Delta\theta)>P(0)} P(\Delta\theta)\,d\Delta\theta$. Here, the assigned probability to identify the tension is $P=\Delta$ \cite{Leizerovich_2024}  and the equivalent $N_\sigma$ is calculated by inverting Eq. \ref{sigma}. Unlike Gaussian tension metrics, which rely on linear approximations and can be highly biased by skewed, non-Gaussian exact parameter shift metric incorporates the full parameter space structure and asymmetric support to produce a conservative and model-independent measurement of tension. In practice, the probability $\Delta$ is estimated numerically from Monte Carlo samples of the shift distribution. Since the estimator is based on a finite number of samples, cases in which no samples fall below the no-shift density correspond to an upper bound on $\Delta$ rather than a vanishing probability. In such cases, we report conservative bounds using a count-based estimator, $\Delta\sim (k+1)/(n+1)$, where $k$ is the number of samples with density lower than $P(0)$ out of $n$ total test samples. This procedure avoids numerical artifacts and ensures a well-defined finite-sample interpretation of the shift significance.

In Tables \ref{tab:lcdm_base_CMB} and \ref{tab:lcdm_full_base_CMB}, it is clearly seen that for the CMB and DESI DR2, the values of the non-Gaussian exact parameter shift and that of $Q^*$ for DM metric is almost similar, indicating very small tension between the dataset combinations. Also, the lower values of DM in $\Lambda_{\rm{s}}$CDM indicate that the model agrees more between shared space datasets. Again, in Tables \ref{tab:lcdm_base_CMB_pp} and \ref{tab:lcdm_full_base_CMB_pp}, in earlier Gaussian and update-based metrics, the apparent tensions involving PPS appear exaggerated due to extreme posterior broadness and covariance compression. However, when applying the exact non-Gaussian parameter shift with finite-sample control, we find that the no-shift point lies in a very low-density region of the shift distribution. Using batched Monte Carlo estimation totaling $5\times10^{6}$ test samples, we find zero samples with density lower than the no-shift point, implying a conservative bound $p_{shift} < 2\times 10^{-7}$, corresponding to $\sigma_{shift}\geq 5.1$. As a result, this indicates that the posteriors occupy disjoint regions of parameter space once their full non-Gaussian structure is accounted for. As an independent Gaussian cross-check, we also compute the Mahalanobis distance \cite{mpc1927, article, DEMAESSCHALCK20001} $d_M$ of the shift distribution and obtain $d_M\sim 5.12$, consistent with a substantial joint displacement in the correlated parameter space. Also, this bound is stable under variations of the KDE bandwidth by a factor of four $0.5-2.0$, indicating that our result is not driven by smoothing choices or numerical artifacts.

To clarify the relationship between Gaussian and exact non-Gaussian tension estimators, we emphasize that agreement or disagreement between these metrics depends sensitively on the geometry of the posterior distributions in the shared parameter space. Gaussian estimators such as DM and UDM approximate the shift between datasets using only the first two moments of the posterior, implicitly assuming local quadratic behavior of the log-likelihood around the posterior mean. This approximation can misrepresent tension when posterior distributions are highly skewed, non-elliptical, truncated by priors, or dominated by one dataset. 

For the CMB+DESI DR2 comparison, both Gaussian and exact non-Gaussian parameter-shift statistics yield consistent results, indicating substantial posterior overlap. This agreement arises because both posteriors are compact and approximately elliptical in the shared parameter subspace, such that the quadratic approximation remains valid. On the other hand, when comparing CMB+DESI DR2 with PPS, Gaussian estimators yield large apparent tensions, and the exact non-Gaussian shift confirms that the no-shift point lies in a region of extremely low probability density. The agreement between Gaussian and exact estimators in this case does not contradict our earlier caution regarding Gaussian metrics. That said, it reflects the fact that the dominant effect is not posterior skewness alone, but a genuine displacement of high-probability support in correlated parameter directions. 


\begin{figure*}[ht]
  \centering
  \includegraphics[width=\linewidth]{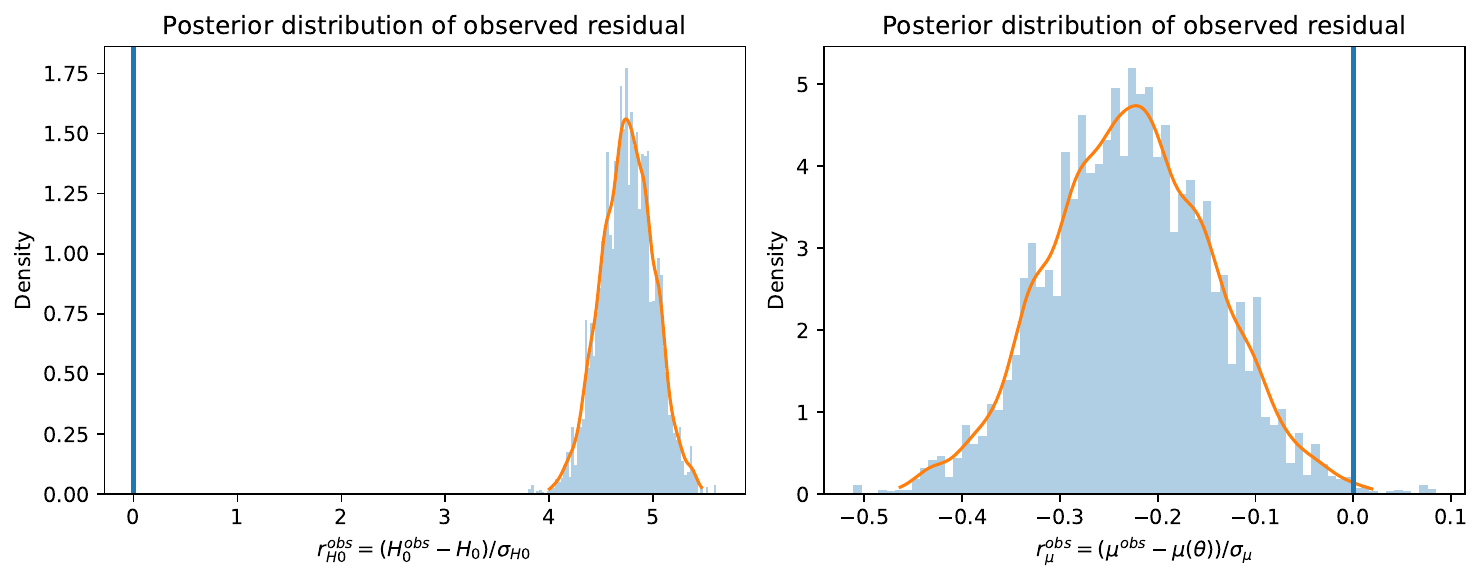}
  \caption{Posterior predictive residual distributions for the $\Lambda$CDM model. Histograms show replicated discrepancies in $H_0$ and $\mu(z=0.01)$ generated from posterior samples constrained by CMB and DESI DR2 data. The vertical line denotes the observed value in each case.}
  \label{fig:lcdm_observed_residuals}
\end{figure*}

\begin{figure}[ht]
  \centering
  \includegraphics[width=1.0\linewidth]{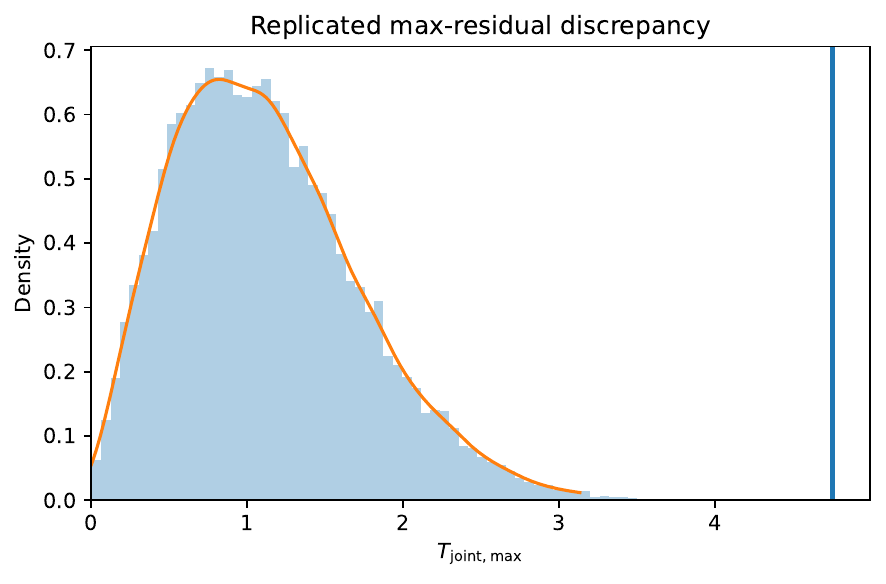}
  \caption{Posterior predictive distribution of the maximum joint discrepancy statistic for $\Lambda$CDM. Replicated values are obtained from posterior draws, and the vertical line indicates the observed statistic.}
  \label{fig:lcdm_max_discrepancy}
\end{figure}

\begin{figure*}[t]
  \centering
  \includegraphics[width=\linewidth]{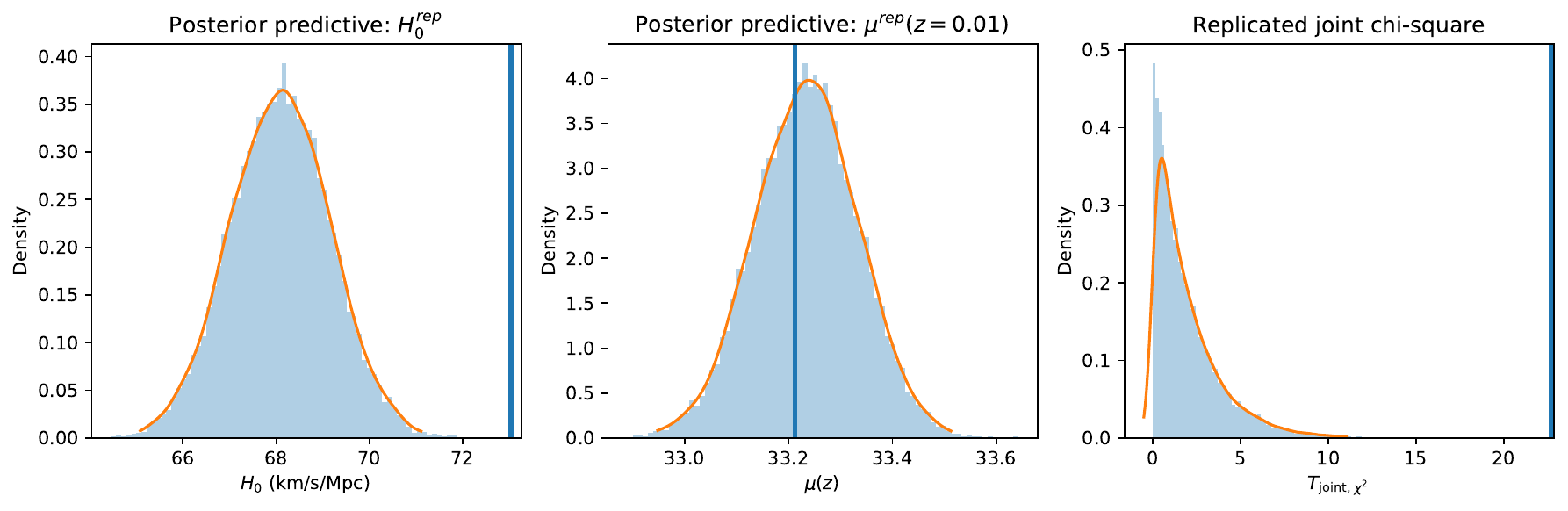}
  \caption{Posterior predictive replicated distributions for $\Lambda$CDM showing $H_0$, $\mu(z=0.01)$, and the joint $\chi^2$ statistic. Replicated values are generated from posterior samples; vertical lines mark the observed quantities.}
  \label{fig:lcdm_replicates}
\end{figure*}

\begin{figure*}[ht]
  \centering
  \includegraphics[width=\linewidth]{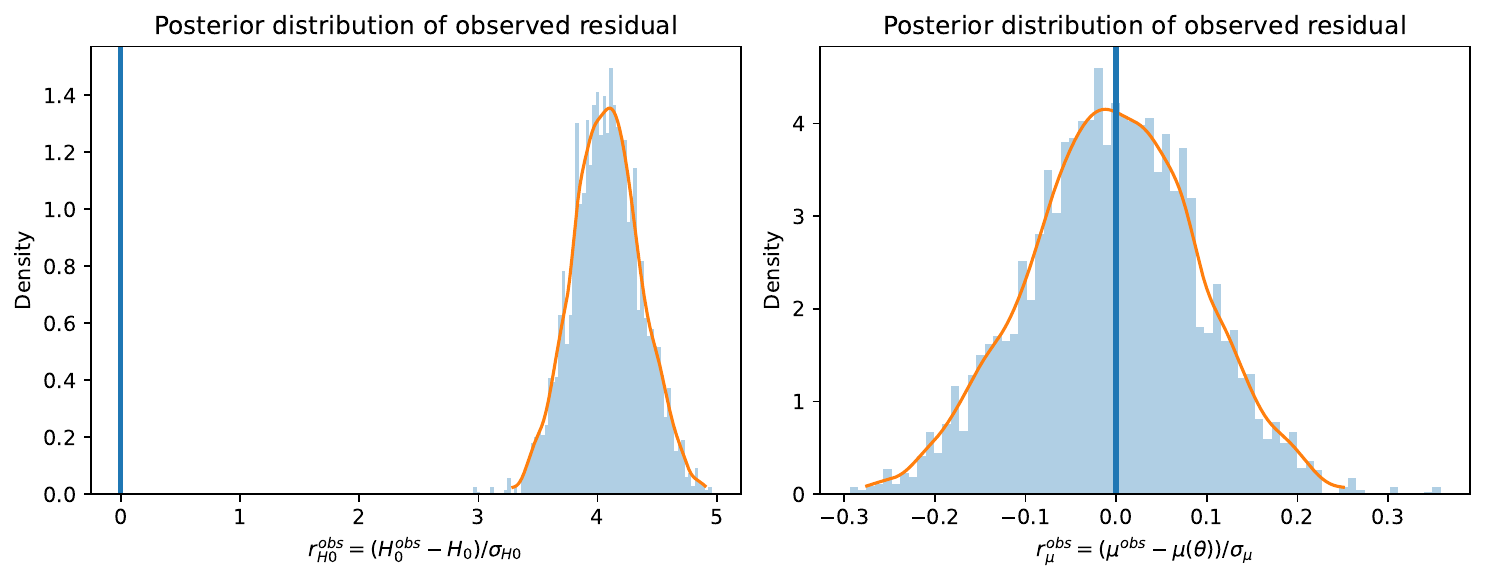}
  \caption{Posterior predictive residual distributions for the $\Lambda_{\rm{s}}$CDM model. Histograms show replicated discrepancies in $H_0$ and $\mu(z=0.01)$ derived from posterior samples constrained by CMB and DESI DR2 data. Vertical lines denote the observed values.}
  \label{fig:lscdm_observed_residuals}
\end{figure*}

\begin{figure*}[t]
  \centering
  \includegraphics[width=\linewidth]{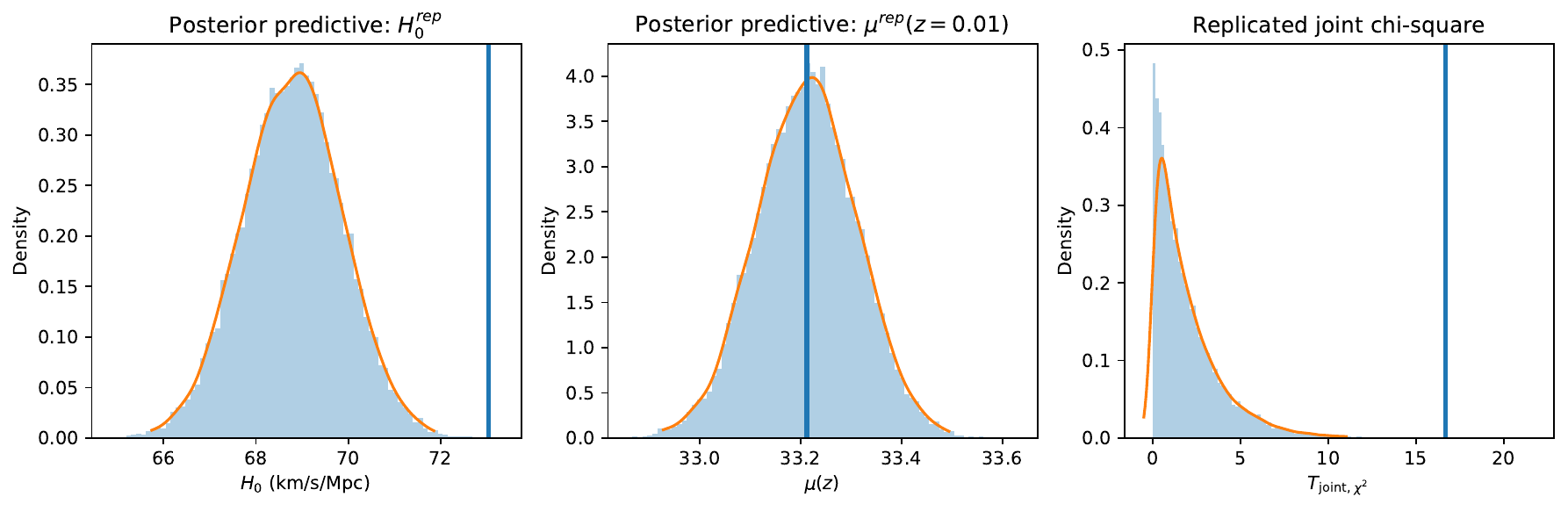}
  \caption{Posterior predictive replicated distributions for $\Lambda_{\rm{s}}$CDM showing $H_0$, $\mu(z=0.01)$, and the joint $\chi^2$ statistic. Replicated samples are generated from the posterior; vertical lines indicate the observed values.}
  \label{fig:lscdm_replicates}
\end{figure*}

\begin{figure}[t]
  \centering
  \includegraphics[width=1.0\linewidth]{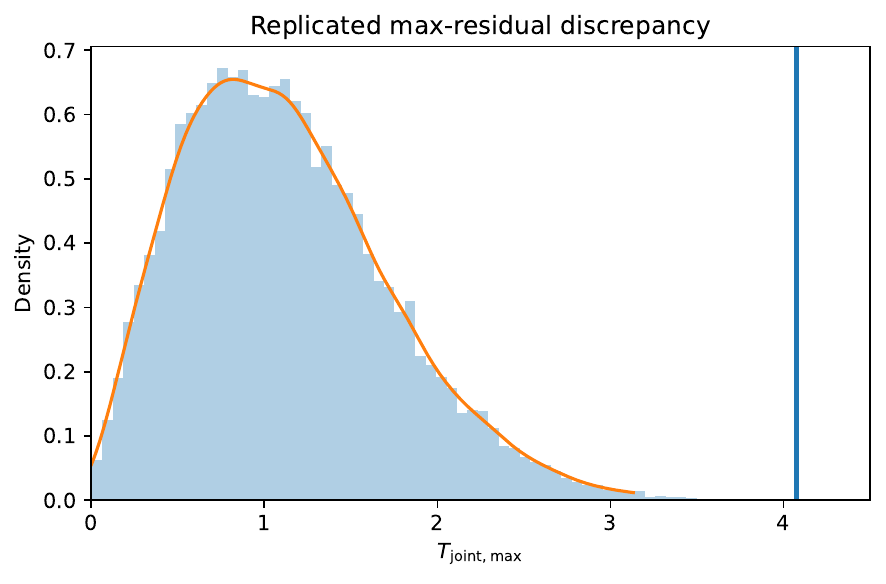}
  \caption{Posterior predictive distribution of the maximum joint discrepancy statistic for $\Lambda_{\rm{s}}$CDM. Replicated values are computed from posterior draws, with the observed statistic indicated by the vertical line.}
  \label{fig:lscdm_max_discrepancy}
\end{figure}
\subsection{Posterior Predictive Consistency Tests}

The posterior predictive distribution describes the probability distribution of future or replicated data conditioned on the observed data through the posterior over model parameters \cite{BSLYNCH2005135, BS2024arXiv241215809S, DES:2020lei}. Formally, it is defined as
\begin{equation}
\label{eq:ppd}
p\left(y^{\mathrm{rep}} \mid y\right)
= \int p\left(y^{\mathrm{rep}} \mid \theta\right)\, p\left(\theta \mid y\right)\, d\theta ,
\end{equation}
where $y^{\mathrm{rep}}$ denotes data that could be generated under the model, $y$ represents the observed data, and $\theta$ is the set of model parameters \cite{BSLYNCH2005135,BS2024arXiv241215809S}. The integrand consists of the sampling distribution for the replicated data conditioned on the parameters and the posterior distribution of the parameters given the observed data. This construction provides a way to assess model adequacy: if a model offers a good description of the observed data, then replicated data drawn from the posterior predictive distribution should resemble the observations. Consequently, model checking can be performed by generating replicated data sets and comparing them to the observed data using suitable discrepancy measures. To quantify this comparison, one may define a test statistic $T(y)$ that captures a relevant feature of the data, and evaluate the same statistic $T(y^{\mathrm{rep}})$ on the replicated data. A Bayesian posterior predictive $p$-value is then given by
\begin{equation}
p_{\mathrm{ppc}} = \Pr\!\left[\, T\!\left(y^{\mathrm{rep}}\right) \geq T(y) \,\middle|\, y \right],
\label{eq:ppc_pvalue}
\end{equation}
which represents the probability that the test statistic computed from replicated data exceeds that obtained from the observed data \cite{Gelman:1996ppc, Meng1994PPP, 2024arXiv241215809S}. Values of $p_{\mathrm{ppc}}$ close to zero or unity indicate a potential mismatch between the model and the observed data, while values near the middle of the unit interval suggest adequate model consistency.

In this work, PPC tests are performed for both the $\Lambda$CDM and $\Lambda_{\rm{s}}$CDM models using posterior samples obtained from CMB and DESI DR2 and large-scale structure data. From each posterior draw, replicated observables are generated for the Hubble constant $H_0$ and a low-redshift supernova distance modulus $\mu(z)$ evaluated at $z=0.01$. To quantify discrepancies, we consider multiple statistics: a one-dimensional statistic for $H_0$, a corresponding statistic for $\mu$, a joint $\chi^2$ statistic constructed from $(H_0,\mu)$, and a maximum joint discrepancy statistic designed to capture worst-case deviations. Posterior predictive $p$-values are computed as the fraction of replicated discrepancies exceeding the observed value, with Monte Carlo uncertainties estimated from repeated draws.

For the $\Lambda$CDM model, the PPC analysis reveals a severe inconsistency with the local measurement of the Hubble constant. The posterior predictive distribution yields a Planck-like expansion rate, and the observed value $H_0 = 73.04 \pm 1.04\,\mathrm{km\,s^{-1}\, Mpc^{-1}}$ \cite{Riess:2021jrx} lies deep in the tail of the predictive distribution refer to Fig. \ref{fig:lcdm_replicates}, resulting in $p_{\mathrm{ppc}}(H_0) \simeq 5\times10^{-5}$. This inconsistency is reinforced by the joint statistics, for which both the joint $\chi^2$ and maximum discrepancy yield $p_{\mathrm{ppc}} \simeq 5\times10^{-5}$. These results indicate that the observed combination of expansion rate and distance information is highly atypical under $\Lambda$CDM. In contrast, the low-redshift supernova distance modulus is well reproduced, refer to Fig. \ref{fig:lcdm_replicates}, with $p_{\mathrm{ppc}}(\mu) \simeq 0.81$, demonstrating that the tension does not originate from supernova distance calibration or luminosity systematics alone.

However, introducing the sign-switching cosmological constant in the $\Lambda_{\rm{s}}$CDM model leads to a modest but systematic shift in the posterior predictive distributions toward higher values of $H_0$. Quantitatively, this manifests as an increase in the joint posterior predictive $p$-value to $p_{\mathrm{ppc}}^{\mathrm{joint}} \simeq 4\times10^{-4}$, indicating partial alleviation of the tension relative to $\Lambda$CDM. Nevertheless, the observed $H_0$ remains in the tail of the predictive distribution, and the joint discrepancy statistics continue to signal a statistically significant inconsistency. Importantly, the $\Lambda_{\rm{s}}$CDM model preserves consistency with the low-redshift supernova distance observable, showing that the extension does not compromise the successful distance predictions of the standard model.

The corresponding PPC Figs. further illustrate this behavior. From Fig. \ref{fig:lcdm_observed_residuals}, it is clear that the $H_0$ observed residual histogram is sitting around a large positive value, while the observed vertical line is far out in the tail. This means that $\Lambda$CDM predicts a lower value of $H_0$ than what is observed. Similarly, in Fig.  \ref{fig:lscdm_observed_residuals}, the observed residual $H_0$ still lies in the far tail, but the histogram peak appears slightly closer to what was observed for $\Lambda$CDM, indicating partial alleviation. Thus, $\Lambda_{\rm{s}}$CDM shifts the late-time behavior enough to move predictions modestly toward the local $H_0$, but the observed $H_0$ remains an outlier.

In Figs. \ref{fig:lcdm_max_discrepancy} and~\ref{fig:lscdm_max_discrepancy}, we can see that in both cases the observed vertical line is far to the right, showing that the data generate a discrepancy far larger than what $\Lambda$CDM  typically predicts. However, for $\Lambda_{\rm{s}}$CDM, the maximum discrepancy is less extreme than in $\Lambda$CDM. This is consistent with some improvement: $\Lambda_{\rm{s}}$CDM reduces the most severe inconsistency but does not remove it.

Again, from the replicated distributions shown in Figs. \ref{fig:lcdm_replicates} and \ref{fig:lscdm_replicates} for the $\Lambda$CDM and $\Lambda_{\rm{s}}$CDM models, it is clearly seen that $H_0^{\rm rep}$ is centred around a Planck-like value, with the observed value lying in the extreme right tail of the distribution. The replicated $\mu^{\mathrm{rep}}$ distribution is tightly centered near the observed value, with the observed line lying well within the bulk. In addition, the replicated joint chi-squared distribution is heavily weighted toward low values, while the observed joint statistic lies far to the right, showing that the combined mismatch is very unlikely under $\Lambda$CDM. Similar behavior is seen in the $\Lambda_{\rm{s}}$CDM replicated distributions; however, the joint chi-squared remains inconsistent, though the discrepancy is visually less stark than in $\Lambda$CDM. Thus, while the replicated distributions for $\mu$ encompass the observed value, the observed $H_0$ and joint discrepancy statistics consistently lie far outside the bulk of the posterior predictive distributions.

Overall, the PPC diagnostics demonstrate that while $\Lambda_{\rm{s}}$CDM softens the Hubble tension compared to $\Lambda$CDM, it does not fully reconcile local expansion-rate measurements with CMB and large-scale-structure-anchored posteriors. The persistence of the inconsistency across multiple discrepancy statistics and its stability against Monte Carlo uncertainty indicate that the residual tension reflects a genuine mismatch in late-time expansion dynamics rather than a statistical fluctuation or posterior-shape artifact. Thus, since $\Lambda_{\rm{s}}$CDM does not effectively decouple the CMB and DESI DR2 ruler calibration in a way that permits a higher $H_0$, it does not make SH0ES + calibrated low-z SN clearly consistent with the CMB+DESI-DR2 posterior, and these PPC results are consistent with the tension-metric analysis: $\Lambda_{\rm{s}}$CDM reduces the severity of the late-time discrepancy relative to $\Lambda$CDM, but the observed $H_0$ (and joint discrepancy statistics) remain extreme under the posterior predictive distribution, while the low-redshift distance-modulus proxy is well reproduced. 

\section{Discussion}
\label{Result}
The statistical analyses presented in Sec. \ref{Stat} show that apparent dataset inconsistency depends sensitively on the diagnostic employed, particularly when combining tightly constrained datasets with weakly informative and strongly non-Gaussian ones. In all configurations considered here, the combined CMB+DESI DR2 dataset yields a compact posterior that tightly constrains the shared parameter space, while the PPS dataset exhibits a substantially larger and highly non-Gaussian posterior. As a result, comparisons involving PPS are inherently asymmetric and require careful interpretation.

Gaussian tension metrics applied to $\Lambda$CDM indicate moderate disagreement between CMB and DESI DR2, and a much stronger discrepancy once PPS is included. When assessed using the exact non-Gaussian parameter-shift statistic, the CMB+DESI DR2 comparison shows substantial posterior overlap, confirming excellent mutual consistency between early and intermediate-redshift probes. In contrast, comparisons involving PPS yield large exact parameter shifts, indicating statistically disjoint posterior support relative to the CMB+DESI DR2 constraints.

When PPS is updated with the highly constrained CMB+DESI DR2 posterior, this effect is more noticeable. Here, the accurate non-Gaussian parameter-shift statistic shows that the PPS posterior fills a specific region of parameter space, whereas updated Gaussian metrics enhance directional mean shifts caused by covariance compression. This shows that PPS-driven Gaussian tensions are not numerical variations of posterior geometry but rather reflect actual late-time inconsistencies, and thus Gaussian metrics alone cannot be used to reliably evaluate datasets with very large and non-Gaussian posteriors.

The $\Lambda_{\rm{s}}$CDM model systematically reduces Gaussian and MAP-based tension measures across all dataset combinations. In particular, the CMB and DESI DR2 comparison exhibits tensions well below the $0.3\sigma$ level, indicating excellent mutual consistency once additional late-time flexibility is introduced. This improvement arises because $\Lambda_{\rm{s}}$CDM allows expansion along degeneracy directions already present in the CMB data, thereby improving geometric compatibility without altering early-universe physics. In comparisons involving PPS, however, large Gaussian update tensions and significant exact non-Gaussian parameter shifts persist, confirming that the apparent inconsistency is driven by conflicting late-time parameter preferences rather than by posterior-shape effects alone. Relative to $\Lambda$CDM, $\Lambda_{\rm{s}}$CDM shows smaller MAP displacements and reduced rule-of-thumb tensions, indicating a statistically improved but not fully consistent joint description of all datasets.

Concerning the posterior predictive consistency test, it provides an independent assessment of model adequacy that is insensitive to posterior geometry alone. For $\Lambda$CDM, the observed local value of $H_0$ lies deep in the tail of the posterior predictive distribution derived from CMB+DESI DR2 data, yielding extremely small predictive $p$-values for both one-dimensional and joint discrepancy statistics. In contrast, the low-redshift distance-modulus proxy is well reproduced, indicating that the dominant tension originates from the late-time expansion rate rather than from supernova distance calibration.

In the $\Lambda_{\rm{s}}$CDM case, the posterior predictive distributions shift modestly but systematically toward higher values of $H_0$, leading to an increase in joint posterior predictive $p$-values and hence a partial alleviation of the discrepancy. However, the joint discrepancy statistics continue to show significant inconsistency, and the observed $H_0$ is still statistically unusual. Importantly, consistency with the low-redshift distance-modulus observable is preserved, demonstrating that the late-time modification does not compromise the successful distance predictions of the standard $\Lambda$CDM framework.

A key insight emerging from the posterior predictive analysis is the distinction between marginal and joint predictive adequacy. The $\Lambda_{\rm{s}}$CDM modification alters the late-time background expansion in a way that preferentially affects one-dimensional observables such as $H_0$, shifting the corresponding posterior predictive distribution toward higher values. However, this localized modification does not sufficiently reorganize the joint structure of expansion-rate and distance predictions required to render the combined $(H_0,\mu)$ discrepancy statistically typical. Consequently, improvements in marginal posterior predictive performance do not translate into full joint consistency. This explains why $\Lambda_{\rm{s}}$CDM systematically softens but does not resolve the Hubble tension. Thus, the model modifies the late-time expansion history enough to shift $H_0$, but not sufficiently to reconcile the global distance ladder implied by CMB and BAO calibrated posterior.

\section{Conclusion}
\label{conclusion}
In this work, we have explored the internal consistency of current cosmological datasets and the late-time predictive performance of the standard $\Lambda$CDM framework and its sign-switching extension, $\Lambda_{\rm{s}}$CDM, by confronting them with a large combination of early, intermediate, and low-redshift observations. Our analysis combines Gaussian and non-Gaussian parameter space diagnostics with posterior predictive consistency tests to robustly evaluate cosmological tensions. It emphasizes that the interpretation of cosmological tensions depends sensitively on the statistical diagnostics employed, particularly when datasets with very different constraining power and posterior geometry are combined.

We find that Gaussian tension metrics can substantially overstate inconsistencies when datasets with very different constraining power and non-Gaussian posteriors are combined. We have shown that the exact non-Gaussian parameter-shift statistics reveal excellent consistency between CMB and BAO datasets in both $\Lambda$CDM and $\Lambda_{\rm{s}}$CDM models, with the $\Lambda_{\rm{s}}$CDM model further improving geometric compatibility at intermediate redshifts. In contrast, comparisons involving PPS data yield large exact parameter shifts in both models, indicating disjoint posterior support to CMB and BAO constraints.

Our posterior predictive tests show that $\Lambda$CDM strongly disfavors the observed local value of the Hubble constant, while $\Lambda_{\rm{s}}$CDM shifts late-time predictions toward higher $H_0$ values and partially alleviates the tension. Nevertheless, the observed expansion rate remains statistically atypical under the predictive distribution, and joint late-time discrepancies still persist. In summary, $\Lambda_{\rm{s}}$CDM improves parameter space compatibility and softens several late-time tensions relative to     $\Lambda$CDM, but does not fully reconcile local expansion-rate measurements with CMB and BAO calibrated posteriors. Hence, our results suggest the necessity of exact, non-Gaussian, and predictive diagnostics for reliably assessing cosmological model consistency, providing a robust benchmark for assessing future cosmological models and datasets. The analysis pipeline followed in this work is flexible and can be extended to a wide range of cosmological models. In a forthcoming work, we will extend this statistical framework to dynamical dark-energy parameterization via energy density \cite{Kumar:2025obb} and other extended cosmological scenarios, to further assess their internal consistency and predictive performance across current cosmological datasets.

\begin{acknowledgments}
S.K. (the first author) acknowledges the financial support and computational facilities provided by Plaksha University.  A.J.S.C.\ acknowledges Conselho Nacional de Desenvolvimento Cient\'{\i}fico e Tecnol\'ogico (CNPq; National Council for Scientific and Technological Development) for partial financial support (Grant No.~305881/2022-1) and Funda\c{c}\~ao da Universidade Federal do Paran\'a (FUNPAR; Paran\'a Federal University Foundation) through public notice 04/2023-Pesquisa/PRPPG/UFPR for partial financial support (Process No.~23075.019406/2023-92), as well as the financial support of the NAPI ``Fen\^omenos Extremos do Universo'' of Funda\c{c}\~ao de Apoio \`a Ci\^encia, Tecnologia e Inova\c{c}\~ao do Paran\'a (NAPI F\'ISICA--FASE~2), under protocol No.~22.687.035-0. S.K. acknowledges the support of the Startup Research Grant from Plaksha University (File No.\ OOR/PU-SRG/2023-24/08). This article is based upon work from the COST Action CA21136 ``Addressing observational tensions in cosmology with systematics and fundamental physics'' (CosmoVerse), supported by COST (European Cooperation in Science and Technology). 
\end{acknowledgments}

\section*{Data Availability}

The observational data and likelihoods used in this work are publicly available from the Planck Legacy Archive and Planck likelihood releases~\cite{Planck:2018vyg,Planck:2018les,Planck:2019cmb}, the ACT DR6 likelihood releases~\cite{ACT:2023kun,AtacamaCosmologyTelescope:2025blo}, the SPT-3G public data products~\cite{SPT:2023jql,SPT-3G:2021eoc,SPT-3G:2022hvq}, the DESI DR2 BAO data release~\cite{DESI:2025dr2lya,DESI:2025dr2bao}, and the PantheonPlus+SH0ES data release~\cite{Scolnic:2021amr}. The modified Boltzmann codes and likelihoods will be made
available upon reasonable request to the authors following the publication of this work.
\newpage

\bibliographystyle{apsrev4-1}
\bibliography{main}

@article{Raveri_2019,
    author = "Raveri, Marco and Hu, Wayne",
    title = "{Concordance and Discordance in Cosmology}",
    eprint = "1806.04649",
    archivePrefix = "arXiv",
    primaryClass = "astro-ph.CO",
    doi = "10.1103/PhysRevD.99.043506",
    journal = "Phys. Rev. D",
    volume = "99",
    number = "4",
    pages = "043506",
    year = "2019"
}

@article{Raveri_2021,
    author = "Raveri, Marco and Doux, Cyrille",
    title = "{Non-Gaussian estimates of tensions in cosmological parameters}",
    eprint = "2105.03324",
    archivePrefix = "arXiv",
    primaryClass = "astro-ph.CO",
    doi = "10.1103/PhysRevD.104.043504",
    journal = "Phys. Rev. D",
    volume = "104",
    number = "4",
    pages = "043504",
    year = "2021"
}

@article{Leizerovich_2024,
    author = "Leizerovich, Matias and Landau, Susana J. and Scoccola, Claudia G.",
    title = "{Tensions in cosmology: A discussion of statistical tools to determine inconsistencies}",
    eprint = "2312.08542",
    archivePrefix = "arXiv",
    primaryClass = "astro-ph.CO",
    doi = "10.1016/j.physletb.2024.138844",
    journal = "Phys. Lett. B",
    volume = "855",
    pages = "138844",
    year = "2024"
}

@article{Raveri_2020,
    author = "Raveri, Marco and Zacharegkas, Georgios and Hu, Wayne",
    title = "{Quantifying concordance of correlated cosmological data sets}",
    eprint = "1912.04880",
    archivePrefix = "arXiv",
    primaryClass = "astro-ph.CO",
    doi = "10.1103/PhysRevD.101.103527",
    journal = "Phys. Rev. D",
    volume = "101",
    number = "10",
    pages = "103527",
    year = "2020"
}

@article{Lemos_2021,
    author = "Lemos, P. and others",
    collaboration = "DES",
    title = "{Assessing tension metrics with dark energy survey and Planck data}",
    eprint = "2012.09554",
    archivePrefix = "arXiv",
    primaryClass = "astro-ph.CO",
    reportNumber = "FERMILAB-PUB-20-662-AE",
    doi = "10.1093/mnras/stab1670",
    journal = "Mon. Not. Roy. Astron. Soc.",
    volume = "505",
    number = "4",
    pages = "6179--6194",
    year = "2021"
}

@article{Planck:2018les,
    author = "Aghanim, N. and others",
    collaboration = "Planck",
    title = "{Planck 2018 results. VIII. Gravitational lensing}",
    eprint = "1807.06210",
    archivePrefix = "arXiv",
    primaryClass = "astro-ph.CO",
    doi = "10.1051/0004-6361/201833886",
    journal = "Astron. Astrophys.",
    volume = "641",
    pages = "A8",
    year = "2020"
}

@article{Planck:2019cmb,
    author = "Aghanim, N. and others",
    collaboration = "Planck",
    title = "{Planck 2018 results. V. CMB power spectra and likelihoods}",
    eprint = "1907.12875",
    archivePrefix = "arXiv",
    primaryClass = "astro-ph.CO",
    doi = "10.1051/0004-6361/201936386",
    journal = "Astron. Astrophys.",
    volume = "641",
    pages = "A5",
    year = "2020"
}

@article{Planck:2018vyg,
    author = "Aghanim, N. and others",
    collaboration = "Planck",
    title = "{Planck 2018 results. VI. Cosmological parameters}",
    eprint = "1807.06209",
    archivePrefix = "arXiv",
    primaryClass = "astro-ph.CO",
    doi = "10.1051/0004-6361/201833910",
    journal = "Astron. Astrophys.",
    volume = "641",
    pages = "A6",
    year = "2020",
    note = "[Erratum: Astron.Astrophys. 652, C4 (2021)]"
}

@article{ACT:2023kun,
    author = "Madhavacheril, Mathew S. and others",
    collaboration = "ACT",
    title = "{The Atacama Cosmology Telescope: DR6 Gravitational Lensing Map and Cosmological Parameters}",
    eprint = "2304.05203",
    archivePrefix = "arXiv",
    primaryClass = "astro-ph.CO",
    reportNumber = "FERMILAB-PUB-23-206-PPD",
    doi = "10.3847/1538-4357/acff5f",
    journal = "Astrophys. J.",
    volume = "962",
    number = "2",
    pages = "113",
    year = "2024"
}

@article{AtacamaCosmologyTelescope:2025blo,
    author = "Louis, Thibaut and others",
    collaboration = "Atacama Cosmology Telescope",
    title = "{The Atacama Cosmology Telescope: DR6 power spectra, likelihoods and {\ensuremath{\Lambda}}CDM parameters}",
    eprint = "2503.14452",
    archivePrefix = "arXiv",
    primaryClass = "astro-ph.CO",
    reportNumber = "FERMILAB-PUB-25-0071-PPD",
    doi = "10.1088/1475-7516/2025/11/062",
    journal = "JCAP",
    volume = "11",
    pages = "062",
    year = "2025"
}

@article{SPT:2023jql,
    author = "Pan, Z. and others",
    collaboration = "SPT",
    title = "{Measurement of gravitational lensing of the cosmic microwave background using SPT-3G 2018 data}",
    eprint = "2308.11608",
    archivePrefix = "arXiv",
    primaryClass = "astro-ph.CO",
    doi = "10.1103/PhysRevD.108.122005",
    journal = "Phys. Rev. D",
    volume = "108",
    number = "12",
    pages = "122005",
    year = "2023"
}

@article{SPT-3G:2021eoc,
    author = "Dutcher, D. and others",
    collaboration = "SPT-3G",
    title = "{Measurements of the E-mode polarization and temperature-E-mode correlation of the CMB from SPT-3G 2018 data}",
    eprint = "2101.01684",
    archivePrefix = "arXiv",
    primaryClass = "astro-ph.CO",
    reportNumber = "FERMILAB-PUB-21-137-AE",
    doi = "10.1103/PhysRevD.104.022003",
    journal = "Phys. Rev. D",
    volume = "104",
    number = "2",
    pages = "022003",
    year = "2021"
}

@article{SPT-3G:2022hvq,
    author = "Balkenhol, L. and others",
    collaboration = "SPT-3G",
    title = "{Measurement of the CMB temperature power spectrum and constraints on cosmology from the SPT-3G 2018 TT, TE, and EE dataset}",
    eprint = "2212.05642",
    archivePrefix = "arXiv",
    primaryClass = "astro-ph.CO",
    reportNumber = "FERMILAB-PUB-22-953-PPD",
    doi = "10.1103/PhysRevD.108.023510",
    journal = "Phys. Rev. D",
    volume = "108",
    number = "2",
    pages = "023510",
    year = "2023"
}

@article{Scolnic:2021amr,
    author = "Scolnic, Dan and others",
    title = "{The Pantheon+ Analysis: The Full Data Set and Light-curve Release}",
    eprint = "2112.03863",
    archivePrefix = "arXiv",
    primaryClass = "astro-ph.CO",
    doi = "10.3847/1538-4357/ac8b7a",
    journal = "Astrophys. J.",
    volume = "938",
    number = "2",
    pages = "113",
    year = "2022"
}

@book{Ryden2017,
    author = "Ryden, B.",
    title = "{Introduction to cosmology}",
    doi = "10.1017/9781316651087",
    isbn = "978-1-107-15483-4, 978-1-316-88984-8, 978-1-316-65108-7",
    publisher = "Cambridge University Press",
    year = "1970"
}

@book{AmendolaTsujikawa2010,
    author = "Amendola, Luca and Tsujikawa, Shinji",
    title = "{Dark Energy}: {Theory and Observations}",
    isbn = "978-1-107-45398-2",
    publisher = "Cambridge University Press",
    month = "1",
    year = "2015"
}

@article{DES:2021wwk,
    author = "Abbott, T. M. C. and others",
    collaboration = "DES",
    title = "{Dark Energy Survey Year 3 results: Cosmological constraints from galaxy clustering and weak lensing}",
    eprint = "2105.13549",
    archivePrefix = "arXiv",
    primaryClass = "astro-ph.CO",
    reportNumber = "FERMILAB-PUB-21-221-AE, DES-2020-0617",
    doi = "10.1103/PhysRevD.105.023520",
    journal = "Phys. Rev. D",
    volume = "105",
    number = "2",
    pages = "023520",
    year = "2022"
}

@article{Akarsu:2019hmw,
    author = {Akarsu, {\"O}zg{\"u}r and Barrow, John D. and Escamilla, Luis A. and Vazquez, J. Alberto},
    title = "{Graduated dark energy: Observational hints of a spontaneous sign switch in the cosmological constant}",
    eprint = "1912.08751",
    archivePrefix = "arXiv",
    primaryClass = "astro-ph.CO",
    doi = "10.1103/PhysRevD.101.063528",
    journal = "Phys. Rev. D",
    volume = "101",
    number = "6",
    pages = "063528",
    year = "2020"
}

@article{Akarsu:2024eoo,
    author = {Akarsu, {O}zg{u}r and De Felice, Antonio and Di Valentino, Eleonora and Kumar, Suresh and Nunes, Rafael C. and {O}z{u}lker, Emre and Vazquez, J. Alberto and Yadav, Anita},
    title = "{Cosmological constraints on LsCDM scenario in a type II minimally modified gravity}",
    eprint = "2406.07526",
    archivePrefix = "arXiv",
    primaryClass = "astro-ph.CO",
    reportNumber = "YITP-24-57",
    doi = "10.1103/PhysRevD.110.103527",
    journal = "Phys. Rev. D",
    volume = "110",
    number = "10",
    pages = "103527",
    year = "2024"
}

@article{Soriano:2025gxd,
    author = "Soriano, Jorge F. and Wohlberg, Shimon and Anchordoqui, Luis A.",
    title = "{New insights on a sign-switching Lambda}",
    eprint = "2502.19239",
    archivePrefix = "arXiv",
    primaryClass = "astro-ph.CO",
    doi = "10.1016/j.dark.2025.101911",
    journal = "Phys. Dark Univ.",
    volume = "48",
    pages = "101911",
    year = "2025"
}

@article{Akarsu:2023mfb,
    author = "Akarsu, Ozgur and Di Valentino, Eleonora and Kumar, Suresh and Nunes, Rafael C. and Vazquez, J. Alberto and Yadav, Anita",
    title = "{$\Lambda_{\rm s}$CDM model: A promising scenario for alleviation of cosmological tensions}",
    eprint = "2307.10899",
    archivePrefix = "arXiv",
    primaryClass = "astro-ph.CO",
    journal =  "arXiv preprint",
    month = "7",
    year = "2023"
}

@article{Akarsu:2021fol,
    author = {Akarsu, \"Ozg\"ur and Kumar, Suresh and \"Oz\"ulker, Emre and Vazquez, J. Alberto},
    title = "{Relaxing cosmological tensions with a sign switching cosmological constant}",
    eprint = "2108.09239",
    archivePrefix = "arXiv",
    primaryClass = "astro-ph.CO",
    doi = "10.1103/PhysRevD.104.123512",
    journal = "Phys. Rev. D",
    volume = "104",
    number = "12",
    pages = "123512",
    year = "2021"
}

@article{escamilla2025exploring,
    author = {Escamilla, Luis A. and Akarsu, {\"O}zg{\"u}r and Di Valentino, Eleonora and {\"O}z{\"u}lker, Emre and Vazquez, J. Alberto},
    title = "{Exploring the Growth-Index ($\gamma$) Tension with $\Lambda_{\rm s}$CDM}",
    eprint = "2503.12945",
    archivePrefix = "arXiv",
    primaryClass = "astro-ph.CO",
    month = "3",
    year = "2025",
    journal ={arXiv}
}

@article{Riess:2025chq,
    author = "Riess, Adam G. and others",
    title = "{The Perfect Host: JWST Cepheid Observations in a Background-free Type Ia Supernova Host Confirm No Bias in Hubble-constant Measurements}",
    eprint = "2509.01667",
    archivePrefix = "arXiv",
    primaryClass = "astro-ph.CO",
    doi = "10.3847/2041-8213/ae0ad6",
    journal = "Astrophys. J. Lett.",
    volume = "992",
    number = "2",
    pages = "L34",
    year = "2025"
}

@article{TDCOSMO:2025dmr,
    author = "Birrer, Simon and others",
    collaboration = "TDCOSMO",
    title = "{TDCOSMO 2025: Cosmological constraints from strong lensing time delays}",
    eprint = "2506.03023",
    archivePrefix = "arXiv",
    primaryClass = "astro-ph.CO",
    reportNumber = "FERMILAB-PUB-25-0381-CSAID",
    doi = "10.1051/0004-6361/202555801",
    journal = "Astron. Astrophys.",
    volume = "704",
    pages = "A63",
    year = "2025"
}

@article{DeFelice:2020eju,
    author = "De Felice, Antonio and Doll, Andreas and Mukohyama, Shinji",
    title = "{A theory of type-II minimally modified gravity}",
    eprint = "2004.12549",
    archivePrefix = "arXiv",
    primaryClass = "gr-qc",
    reportNumber = "YITP-20-55, IPMU20-0040",
    doi = "10.1088/1475-7516/2020/09/034",
    journal = "JCAP",
    volume = "09",
    pages = "034",
    year = "2020"
}

@article{PhysRevD.109.103522,
  title = {Transition dynamics in the ${\mathrm{\ensuremath{\Lambda}}}_{\mathrm{s}}\mathrm{CDM}$ model: Implications for bound cosmic structures},
  author = {Paraskevas, Evangelos A. and \ifmmode \mbox{\c{C}}\else \c{C}\fi{}am, Arman and Perivolaropoulos, Leandros and Akarsu, \"Ozg\"ur},
  journal = {Phys. Rev. D},
  volume = {109},
  issue = {10},
  pages = {103522},
  numpages = {23},
  year = {2024},
  month = {May},
  publisher = {American Physical Society},
  doi = {10.1103/PhysRevD.109.103522},
  url = {https://link.aps.org/doi/10.1103/PhysRevD.109.103522}
}

@article{Bousso:2000xa,
    author = "Bousso, Raphael and Polchinski, Joseph",
    title = "{Quantization of four form fluxes and dynamical neutralization of the cosmological constant}",
    eprint = "hep-th/0004134",
    archivePrefix = "arXiv",
    reportNumber = "SU-ITP-00-12, NSF-ITP-00-40",
    doi = "10.1088/1126-6708/2000/06/006",
    journal = "JHEP",
    volume = "06",
    pages = "006",
    year = "2000"
}

@book{Weinberg:2008zzc,
    author = "Weinberg, Steven",
    title = "{Cosmology}",
    publisher = "{Oxford University Press}",
    isbn = "978-0-19-852682-7",
    year = "2008"
}

@book{Peebles:1994xt,
    author = "Peebles, P. J. E.",
    title = "{Principles of Physical Cosmology}",
    isbn = "978-0-691-20981-4",
    publisher = "Princeton University Press",
    month = "9",
    year = "2020"
}

@book{Carroll:2004st,
    author = "Carroll, Sean M.",
    title = "{Spacetime and Geometry}: {An Introduction to General Relativity}",
    doi = "10.1017/9781108770385",
    isbn = "978-0-8053-8732-2, 978-1-108-48839-6, 978-1-108-77555-7",
    publisher = "Cambridge University Press",
    month = "7",
    year = "2019"
}

@article{riess1998observational,
    author = "Riess, Adam G. and others",
    collaboration = "Supernova Search Team",
    title = "{Observational evidence from supernovae for an accelerating universe and a cosmological constant}",
    eprint = "astro-ph/9805201",
    archivePrefix = "arXiv",
    doi = "10.1086/300499",
    journal = "Astron. J.",
    volume = "116",
    pages = "1009--1038",
    year = "1998"
}

@article{perlmutter1999measurements,
    author = "Perlmutter, S. and others",
    collaboration = "Supernova Cosmology Project",
    title = "{Measurements of $\Omega$ and $\Lambda$ from 42 High Redshift Supernovae}",
    eprint = "astro-ph/9812133",
    archivePrefix = "arXiv",
    reportNumber = "LBNL-41801, LBL-41801",
    doi = "10.1086/307221",
    journal = "Astrophys. J.",
    volume = "517",
    pages = "565--586",
    year = "1999"
}

@ARTICLE{2013PhR...530...87W,
       author = {{Weinberg}, David H. and {Mortonson}, Michael J. and {Eisenstein}, Daniel J. and {Hirata}, Christopher and {Riess}, Adam G. and {Rozo}, Eduardo},
        title = "{Observational probes of cosmic acceleration}",
      journal = {physrep},
         year = 2013,
        month = sep,
       volume = {530},
       number = {2},
        pages = {87-255},
          doi = {10.1016/j.physrep.2013.05.001},
archivePrefix = {arXiv},
       eprint = {1201.2434},
 primaryClass = {astro-ph.CO},
       adsurl = {https://ui.adsabs.harvard.edu/abs/2013PhR...530...87W}
}

@article{WMAP:2003elm,
    author = "Spergel, D. N. and others",
    collaboration = "WMAP",
    title = "{First year Wilkinson Microwave Anisotropy Probe (WMAP) observations: Determination of cosmological parameters}",
    eprint = "astro-ph/0302209",
    archivePrefix = "arXiv",
    doi = "10.1086/377226",
    journal = "Astrophys. J. Suppl.",
    volume = "148",
    pages = "175--194",
    year = "2003"
}

@article{percival2007measuring,
    author = {Percival, Will J. and Cole, Shaun and others},
    title = {Measuring the Baryon Acoustic Oscillation scale using the Sloan Digital Sky Survey and 2dF Galaxy Redshift Survey},
    journal = {MNRAS},
    volume = {381},
    number = {3},
    pages = {1053-1066},
    year = {2007},
    month = {10},
    issn = {0035-8711},
    doi = {10.1111/j.1365-2966.2007.12268.x},
    url = {https://doi.org/10.1111/j.1365-2966.2007.12268.x},
}

@article{alam2021completed,
    author = "Alam, Shadab and others",
    collaboration = "eBOSS",
    title = "{Completed SDSS-IV extended Baryon Oscillation Spectroscopic Survey: Cosmological implications from two decades of spectroscopic surveys at the Apache Point Observatory}",
    eprint = "2007.08991",
    archivePrefix = "arXiv",
    primaryClass = "astro-ph.CO",
    doi = "10.1103/PhysRevD.103.083533",
    journal = "Phys. Rev. D",
    volume = "103",
    number = "8",
    pages = "083533",
    year = "2021"
}

@ARTICLE{komatsu,
    author = {{Komatsu}, E. and {Smith}, K.~M. and {Dunkley}, others},
    title = "{Seven-year Wilkinson Microwave Anisotropy Probe (WMAP) Observations: Cosmological Interpretation}",
    journal = {apjs},
    year = {2011},
    month = {feb},
    volume = {192},
    number = {2},
    eid = {18},
    pages = {18},
    doi = {10.1088/0067-0049/192/2/18},
    archivePrefix = {arXiv},
    eprint = {1001.4538},
    primaryClass = {astro-ph.CO},
    adsurl = {https://ui.adsabs.harvard.edu/abs/2011ApJS..192...18K}
}

@article{madhavacheril,
doi = {10.3847/1538-4357/acff5f},
url = {https://doi.org/10.3847/1538-4357/acff5f},
year = {2024},
month = {feb},
publisher = {The American Astronomical Society},
volume = {962},
number = {2},
pages = {113},
author = {Madhavacheril, Mathew S. and others},
title = {The Atacama Cosmology Telescope: DR6 Gravitational Lensing Map and Cosmological Parameters},
journal = {The Astrophysical Journal}
}

@ARTICLE{beutler,
       author = {{Beutler}, Florian and {Blake}, others},
        title = "{The 6dF Galaxy Survey: baryon acoustic oscillations and the local Hubble constant}",
      journal = {mnras},
         year = 2011,
        month = oct,
       volume = {416},
       number = {4},
        pages = {3017-3032},
          doi = {10.1111/j.1365-2966.2011.19250.x},
archivePrefix = {arXiv},
       eprint = {1106.3366},
 primaryClass = {astro-ph.CO},
       adsurl = {https://ui.adsabs.harvard.edu/abs/2011MNRAS.416.3017B}
}

@ARTICLE{parkinson,
       author = {{Parkinson}, David and {Riemer-S{o}rensen}, others},
        title = "{The WiggleZ Dark Energy Survey: Final data release and cosmological results}",
      journal = {prd},
         year = 2012,
        month = nov,
       volume = {86},
       number = {10},
          eid = {103518},
        pages = {103518},
          doi = {10.1103/PhysRevD.86.103518},
archivePrefix = {arXiv},
       eprint = {1210.2130},
 primaryClass = {astro-ph.CO},
       adsurl = {https://ui.adsabs.harvard.edu/abs/2012PhRvD..86j3518P}
}

@ARTICLE{Alam2017,
       author = {{Alam}, Shadab and {Ata} and others},
        title = "{The clustering of galaxies in the completed SDSS-III Baryon Oscillation Spectroscopic Survey: cosmological analysis of the DR12 galaxy sample}",
      journal = {\mnras},
         year = 2017,
        month = sep,
       volume = {470},
       number = {3},
        pages = {2617-2652},
          doi = {10.1093/mnras/stx721},
archivePrefix = {arXiv},
       eprint = {1607.03155},
 primaryClass = {astro-ph.CO},
       adsurl = {https://ui.adsabs.harvard.edu/abs/2017MNRAS.470.2617A}
}

@article{bartelmann2001weak,
    author = "Bartelmann, Matthias and Schneider, Peter",
    title = "{Weak gravitational lensing}",
    eprint = "astro-ph/9912508",
    archivePrefix = "arXiv",
    doi = "10.1016/S0370-1573(00)00082-X",
    journal = "Phys. Rept.",
    volume = "340",
    pages = "291--472",
    year = "2001"
}

@ARTICLE{2015Liu,
    author = {{Liu}, Xiangkun and {Pan}, Chuzhong and {Li}, others},
    title = "{Cosmological constraints from weak lensing peak statistics with Canada-France-Hawaii Telescope Stripe 82 Survey}",
    journal = {\mnras},
     year = 2015,
    month = jul,
    volume = {450},
    number = {3},
    pages = {2888-2902},
    doi = {10.1093/mnras/stv784},
    archivePrefix = {arXiv},
    eprint = {1412.3683},
    primaryClass = {astro-ph.CO},
    adsurl = {https://ui.adsabs.harvard.edu/abs/2015MNRAS.450.2888L}
}

@ARTICLE{2018Shan,
    author = {{Shan}, HuanYuan and {Liu}, Xiangkun and {Hildebrandt}, others},
    title = "{KiDS-450: cosmological constraints from weak lensing peak statistics - I. Inference from analytical prediction of high signal-to-noise ratio convergence peaks}",
    journal = {\mnras},
    year = 2018,
    month = feb,
    volume = {474},
    number = {1},
    pages = {1116-1134},
    doi = {10.1093/mnras/stx2837},
    archivePrefix = {arXiv},
    eprint = {1709.07651},
    primaryClass = {astro-ph.CO},
    adsurl = {https://ui.adsabs.harvard.edu/abs/2018MNRAS.474.1116S}
}

@ARTICLE{2018Mandelbaum,
       author = {{Mandelbaum}, Rachel},
        title = "{Weak Lensing for Precision Cosmology}",
      journal = {ARAA},
         year = 2018,
        month = sep,
       volume = {56},
        pages = {393-433},
          doi = {10.1146/annurev-astro-081817-051928},
archivePrefix = {arXiv},
       eprint = {1710.03235},
 primaryClass = {astro-ph.CO},
       adsurl = {https://ui.adsabs.harvard.edu/abs/2018ARA&A..56..393M}
}

@article{Heymans:2020gsg,
    author = "Heymans, Catherine and others",
    title = "{KiDS-1000 Cosmology: Multi-probe weak gravitational lensing and spectroscopic galaxy clustering constraints}",
    eprint = "2007.15632",
    archivePrefix = "arXiv",
    primaryClass = "astro-ph.CO",
    doi = "10.1051/0004-6361/202039063",
    journal = "Astron. Astrophys.",
    volume = "646",
    pages = "A140",
    year = "2021"
}

@ARTICLE{Ratra,
       author = {{Peebles}, P.~J. and {Ratra}, Bharat},
        title = "{The cosmological constant and dark energy}",
      journal = {Reviews of Modern Physics},
         year = 2003,
        month = apr,
       volume = {75},
       number = {2},
        pages = {559-606},
          doi = {10.1103/RevModPhys.75.559},
archivePrefix = {arXiv},
       eprint = {astro-ph/0207347},
 primaryClass = {astro-ph},
       adsurl = {https://ui.adsabs.harvard.edu/abs/2003RvMP...75..559P}
}

@article{CosmoVerseNetwork:2025alb,
    author = "Di Valentino, Eleonora and others",
    collaboration = "CosmoVerse Network",
    title = "{The CosmoVerse White Paper: Addressing observational tensions in cosmology with systematics and fundamental physics}",
    eprint = "2504.01669",
    archivePrefix = "arXiv",
    primaryClass = "astro-ph.CO",
    doi = "10.1016/j.dark.2025.101965",
    journal = "Phys. Dark Univ.",
    volume = "49",
    pages = "101965",
    year = "2025"
}

@ARTICLE{2022ApJ...934L...7R,
    author = {{Riess}, Adam G. and {Yuan}, Wenlong and {Macri}, others},
    title = "{A Comprehensive Measurement of the Local Value of the Hubble Constant with 1 km s$^{-1}$ Mpc$^{-1}$ Uncertainty from the Hubble Space Telescope and the SH0ES Team}",
    journal = {{ApJL}},
    year = 2022,
    month = jul,
    volume = {934},
    number = {1},
    eid = {L7},
    pages = {L7},
    doi = {10.3847/2041-8213/ac5c5b},
    archivePrefix = {arXiv},
    eprint = {2112.04510},
    primaryClass = {astro-ph.CO},
    adsurl = {https://ui.adsabs.harvard.edu/abs/2022ApJ...934L...7R}
}

@article{Freedman:2024eph,
    author = "Freedman, Wendy L. and Madore, Barry F. and Hoyt, Taylor and others.",
    title = "{Status Report on the Chicago-Carnegie Hubble Program (CCHP): Measurement of the Hubble Constant Using the Hubble and James Webb Space Telescopes}",
    eprint = "2408.06153",
    archivePrefix = "arXiv",
    primaryClass = "astro-ph.CO",
    doi = "10.3847/1538-4357/adce78",
    journal = "Astrophys. J.",
    volume = "985",
    number = "2",
    pages = "203",
    year = "2025",
    note = "[Erratum: Astrophys.J. 993, 252 (2025)]"
}

@article{Freedman:2019jwv,
    author = "Freedman, Wendy L. and others",
    title = "{The Carnegie-Chicago Hubble Program. VIII. An Independent Determination of the Hubble Constant Based on the Tip of the Red Giant Branch}",
    eprint = "1907.05922",
    archivePrefix = "arXiv",
    primaryClass = "astro-ph.CO",
    doi = "10.3847/1538-4357/ab2f73",
    journal = "Astrophys. J.",
    volume = "882",
    pages = "34",
    year = "2019"
}

@article{H0DN:2025lyy,
    author = "Casertano, Stefano and others",
    collaboration = "H0DN",
    title = "{The Local Distance Network: a community consensus report on the measurement of the Hubble constant at 1{\%} precision}",
    eprint = "2510.23823",
    archivePrefix = "arXiv",
    journal = "arXiv preprint",
    primaryClass = "astro-ph.CO",
    month = "10",
    year = "2025"
}

@article{Verde:2019ivm,
    author = "Verde, L. and Treu, T. and Riess, A. G.",
    title = "{Tensions between the Early and the Late Universe}",
    eprint = "1907.10625",
    archivePrefix = "arXiv",
    primaryClass = "astro-ph.CO",
    doi = "10.1038/s41550-019-0902-0",
    journal = "Nature Astron.",
    volume = "3",
    pages = "891",
    year = "2019"
}

@article{Abdalla:2022yfr,
    author = "Abdalla, Elcio and others",
    title = "{Cosmology intertwined: A review of the particle physics, astrophysics, and cosmology associated with the cosmological tensions and anomalies}",
    eprint = "2203.06142",
    archivePrefix = "arXiv",
    primaryClass = "astro-ph.CO",
    reportNumber = "FERMILAB-CONF-22-192-SCD",
    doi = "10.1016/j.jheap.2022.04.002",
    journal = "JHEAp",
    volume = "34",
    pages = "49--211",
    year = "2022"
}

@article{DiValentino:2021izs,
    author = "Di Valentino, Eleonora and Mena, Olga and Pan, Supriya and Visinelli, Luca and Yang, Weiqiang and Melchiorri, Alessandro and Mota, David F. and Riess, Adam G. and Silk, Joseph",
    title = "{In the realm of the Hubble tension{\textemdash}a review of solutions}",
    eprint = "2103.01183",
    archivePrefix = "arXiv",
    primaryClass = "astro-ph.CO",
    reportNumber = "IPPP/20/108",
    doi = "10.1088/1361-6382/ac086d",
    journal = "Class. Quant. Grav.",
    volume = "38",
    number = "15",
    pages = "153001",
    year = "2021"
}

@article{Feeney:2017sgx,
    author = "Feeney, Stephen M. and Mortlock, Daniel J. and Dalmasso, Niccol{\`o}",
    title = "{Clarifying the Hubble constant tension with a Bayesian hierarchical model of the local distance ladder}",
    eprint = "1707.00007",
    archivePrefix = "arXiv",
    primaryClass = "astro-ph.CO",
    doi = "10.1093/mnras/sty418",
    journal = "Mon. Not. Roy. Astron. Soc.",
    volume = "476",
    number = "3",
    pages = "3861--3882",
    year = "2018"
}

@incollection{BSLYNCH2005135,
    title = {Bayesian Statistics},
    editor = {Kimberly Kempf-Leonard},
    booktitle = {Encyclopedia of Social Measurement},
    publisher = {Elsevier},
    address = {New York},
    pages = {135-144},
    year = {2005},
    isbn = {978-0-12-369398-3},
    doi = {https://doi.org/10.1016/B0-12-369398-5/00156-0},
    url = {https://www.sciencedirect.com/science/article/pii/B0123693985001560},
    author = {Scott M. Lynch}
}

@ARTICLE{BS2024arXiv241215809S,
       author = {{Sennhenn-Reulen}, Holger},
        title = "{Prior-Posterior Derived-Predictive Consistency Checks for Post-Estimation Calculated Quantities of Interest (QOI-Check)}",
      journal = {arXiv e-prints},
         year = 2024,
        month = dec,
          eid = {arXiv:2412.15809},
        pages = {arXiv:2412.15809},
          doi = {10.48550/arXiv.2412.15809},
archivePrefix = {arXiv},
       eprint = {2412.15809},
 primaryClass = {stat.ME},
       adsurl = {https://ui.adsabs.harvard.edu/abs/2024arXiv241215809S}
}

@ARTICLE{CLASS,
       author = {{Lesgourgues}, Julien},
        title = "{The Cosmic Linear Anisotropy Solving System (CLASS) I: Overview}",
      journal = {arXiv e-prints},
         year = 2011,
        month = apr,
          eid = {arXiv:1104.2932},
        pages = {arXiv:1104.2932},
          doi = {10.48550/arXiv.1104.2932},
archivePrefix = {arXiv},
       eprint = {1104.2932},
 primaryClass = {astro-ph.IM},
       adsurl = {https://ui.adsabs.harvard.edu/abs/2011arXiv1104.2932L}
}

@ARTICLE{MH2015arXiv150401896R,
       author = {{Robert}, Christian P.},
        title = "{The Metropolis-Hastings algorithm}",
      journal = {arXiv e-prints},
         year = 2015,
        month = apr,
          eid = {arXiv:1504.01896},
        pages = {arXiv:1504.01896},
          doi = {10.48550/arXiv.1504.01896},
archivePrefix = {arXiv},
       eprint = {1504.01896},
 primaryClass = {stat.CO},
       adsurl = {https://ui.adsabs.harvard.edu/abs/2015arXiv150401896R}
}

@ARTICLE{Gellmaan_R,
       author = {{Vats}, Dootika and {Knudson}, Christina},
        title = "{Revisiting the Gelman-Rubin Diagnostic}",
      journal = {arXiv e-prints},
         year = 2018,
        month = dec,
          eid = {arXiv:1812.09384},
        pages = {arXiv:1812.09384},
          doi = {10.48550/arXiv.1812.09384},
archivePrefix = {arXiv},
       eprint = {1812.09384},
 primaryClass = {stat.CO},
       adsurl = {https://ui.adsabs.harvard.edu/abs/2018arXiv181209384V}
}

@article{Leizerovich:2023qqt,
    author = "Leizerovich, Mat{\'\i}as and Landau, Susana J. and Sc{\'o}ccola, Claudia G.",
    title = "{Tensions in cosmology: A discussion of statistical tools to determine inconsistencies}",
    eprint = "2312.08542",
    archivePrefix = "arXiv",
    primaryClass = "astro-ph.CO",
    doi = "10.1016/j.physletb.2024.138844",
    journal = "Phys. Lett. B",
    volume = "855",
    pages = "138844",
    year = "2024"
}

@ARTICLE{2019arXiv191201134S,
       author = {{Staudte}, Robert G.},
        title = "{Evidence for goodness of fit in Karl Pearson chi-squared statistics}",
      journal = {arXiv e-prints},
         year = 2019,
        month = dec,
          eid = {arXiv:1912.01134},
        pages = {arXiv:1912.01134},
          doi = {10.48550/arXiv.1912.01134},
archivePrefix = {arXiv},
       eprint = {1912.01134},
 primaryClass = {math.ST},
       adsurl = {https://ui.adsabs.harvard.edu/abs/2019arXiv191201134S}
}

@article{DES:2020hen,
    author = "Lemos, P. and others",
    collaboration = "DES",
    title = "{Assessing tension metrics with dark energy survey and Planck data}",
    eprint = "2012.09554",
    archivePrefix = "arXiv",
    primaryClass = "astro-ph.CO",
    reportNumber = "FERMILAB-PUB-20-662-AE",
    doi = "10.1093/mnras/stab1670",
    journal = "Mon. Not. Roy. Astron. Soc.",
    volume = "505",
    number = "4",
    pages = "6179--6194",
    year = "2021"
}

@ARTICLE{2019NatAs...3..891V,
       author = {{Verde}, Licia and {Treu}, Tommaso and {Riess}, Adam G.},
        title = "{Tensions between the early and late Universe}",
      journal = {Nature Astronomy},
         year = 2019,
        month = sep,
       volume = {3},
        pages = {891-895},
          doi = {10.1038/s41550-019-0902-0},
archivePrefix = {arXiv},
       eprint = {1907.10625},
 primaryClass = {astro-ph.CO},
       adsurl = {https://ui.adsabs.harvard.edu/abs/2019NatAs...3..891V}
}

@article{Freedman:2021ahq,
    author = "Freedman, Wendy L.",
    title = "{Measurements of the Hubble Constant: Tensions in Perspective}",
    eprint = "2106.15656",
    archivePrefix = "arXiv",
    primaryClass = "astro-ph.CO",
    doi = "10.3847/1538-4357/ac0e95",
    journal = "Astrophys. J.",
    volume = "919",
    number = "1",
    pages = "16",
    year = "2021"
}

@article{PhysRevD.100.043504,
  title = {Quantifying tensions in cosmological parameters: Interpreting the DES evidence ratio},
  author = {Handley, Will and Lemos, Pablo},
  journal = {Phys. Rev. D},
  volume = {100},
  issue = {4},
  pages = {043504},
  numpages = {15},
  year = {2019},
  month = {Aug},
  publisher = {American Physical Society},
  doi = {10.1103/PhysRevD.100.043504},
  url = {https://link.aps.org/doi/10.1103/PhysRevD.100.043504}
}

@article{DES:2020lei,
    author = "Doux, C. and others",
    collaboration = "DES",
    title = "{Dark energy survey internal consistency tests of the joint cosmological probes analysis with posterior predictive distributions}",
    eprint = "2011.03410",
    archivePrefix = "arXiv",
    primaryClass = "astro-ph.CO",
    reportNumber = "FERMILAB-PUB-20-547-AE",
    doi = "10.1093/mnras/stab526",
    journal = "Mon. Not. Roy. Astron. Soc.",
    volume = "503",
    number = "2",
    pages = "2688--2705",
    year = "2021"
}

@article{Meng1994PPP,
  title   = {Posterior predictive p-values},
  author  = {Meng, Xiao-Li},
  journal = {The Annals of Statistics},
  volume  = {22},
  number  = {3},
  pages   = {1142--1160},
  year    = {1994},
  doi     = {10.1214/aos/1176325622}
}

@ARTICLE{2024arXiv241215809S,
       author = {{Sennhenn-Reulen}, Holger},
        title = "{Prior-Posterior Derived-Predictive Consistency Checks for Post-Estimation Calculated Quantities of Interest (QOI-Check)}",
      journal = {arXiv e-prints},
         year = 2024,
        month = dec,
          eid = {arXiv:2412.15809},
        pages = {arXiv:2412.15809},
          doi = {10.48550/arXiv.2412.15809},
archivePrefix = {arXiv},
       eprint = {2412.15809},
 primaryClass = {stat.ME},
       adsurl = {https://ui.adsabs.harvard.edu/abs/2024arXiv241215809S}
}

@article{mpc1927,
  title={Analysis of race mixture in Bengal},
  author={Mahalanobis, P. C.},
  journal={J. Proc. Asiatic Soc. Bengal},
  volume={23},
  pages={301-333},
  year={1927},
}

@article{article,
author = {Mahalanobis, P.C.},
year = {1936},
month = {01},
pages = {49-55.},
title = {On the Generalized Distance in Statistics},
volume = {12},
journal = {Proc. Natl. Inst. Sci. India}
}

@article{DEMAESSCHALCK20001,
title = {The Mahalanobis distance},
journal = {Chemometr. Intell. Lab. Syst.},
volume = {50},
number = {1},
pages = {1-18},
year = {2000},
issn = {0169-7439},
doi = {https://doi.org/10.1016/S0169-7439(99)00047-7},
url = {https://www.sciencedirect.com/science/article/pii/S0169743999000477},
author = {R. {De Maesschalck} and D. Jouan-Rimbaud and D.L. Massart},
}

@article{Mortonson:2013zfa,
    author = "Mortonson, Michael J. and Weinberg, David H. and White, Martin",
    title = "{Dark Energy: A Short Review}",
    eprint = "1401.0046",
    archivePrefix = "arXiv",
    primaryClass = "astro-ph.CO",
    month = "12",
    year = "2013",
    journal = "arxiv-print"
}

@article{Huterer:2017buf,
    author = "Huterer, Dragan and Shafer, Daniel L",
    title = "{Dark energy two decades after: Observables, probes, consistency tests}",
    eprint = "1709.01091",
    archivePrefix = "arXiv",
    primaryClass = "astro-ph.CO",
    doi = "10.1088/1361-6633/aa997e",
    journal = "Rept. Prog. Phys.",
    volume = "81",
    number = "1",
    pages = "016901",
    year = "2018"
}

@article{Spiegelhalter:2002yvw,
    author = "Spiegelhalter, David J. and Best, Nicola G. and Carlin, Bradley P. and van der Linde, Angelika",
    title = "{Bayesian measures of model complexity and fit}",
    doi = "10.1111/1467-9868.00353",
    journal = "J. Roy. Statist. Soc. B",
    volume = "64",
    number = "4",
    pages = "583--639",
    year = "2002"
}

@article{Riess:2021jrx,
    author = "Riess, Adam G. and others",
    title = "{A Comprehensive Measurement of the Local Value of the Hubble Constant with 1 km s$^{−1}$ Mpc$^{−1}$ Uncertainty from the Hubble Space Telescope and the SH0ES Team}",
    eprint = "2112.04510",
    archivePrefix = "arXiv",
    primaryClass = "astro-ph.CO",
    doi = "10.3847/2041-8213/ac5c5b",
    journal = "Astrophys. J. Lett.",
    volume = "934",
    number = "1",
    pages = "L7",
    year = "2022"
}

@article{Kamionkowski:2022pkx,
    author = "Kamionkowski, Marc and Riess, Adam G.",
    title = "{The Hubble Tension and Early Dark Energy}",
    eprint = "2211.04492",
    archivePrefix = "arXiv",
    primaryClass = "astro-ph.CO",
    doi = "10.1146/annurev-nucl-111422-024107",
    journal = "Ann. Rev. Nucl. Part. Sci.",
    volume = "73",
    pages = "153--180",
    year = "2023"
}

@article{Najafi:2024qzm,
    author = "Najafi, Mahdi and Pan, Supriya and Di Valentino, Eleonora and Firouzjaee, Javad T.",
    title = "{Dynamical dark energy confronted with multiple CMB missions}",
    eprint = "2407.14939",
    archivePrefix = "arXiv",
    primaryClass = "astro-ph.CO",
    doi = "10.1016/j.dark.2024.101539",
    journal = "Phys. Dark Univ.",
    volume = "45",
    pages = "101539",
    year = "2024"
}

@book{CANTATA:2021asi,
    author = "Akrami, Yashar and others",
    collaboration = "CANTATA",
    title = "{Modified Gravity and Cosmology. An Update by the CANTATA Network}",
    eprint = "2105.12582",
    archivePrefix = "arXiv",
    primaryClass = "gr-qc",
    doi = "10.1007/978-3-030-83715-0",
    isbn = "978-3-030-83714-3, 978-3-030-83717-4, 978-3-030-83715-0",
    publisher = "Springer",
    year = "2021",
    journal={arXiv}
}

@article{Sotiriou:2008rp,
    author = "Sotiriou, Thomas P. and Faraoni, Valerio",
    title = "{f(R) Theories Of Gravity}",
    eprint = "0805.1726",
    archivePrefix = "arXiv",
    primaryClass = "gr-qc",
    doi = "10.1103/RevModPhys.82.451",
    journal = "Rev. Mod. Phys.",
    volume = "82",
    pages = "451--497",
    year = "2010"
}

@article{Ma:1995ey,
    author = "Ma, Chung-Pei and Bertschinger, Edmund",
    title = "{Cosmological perturbation theory in the synchronous and conformal Newtonian gauges}",
    eprint = "astro-ph/9506072",
    archivePrefix = "arXiv",
    doi = "10.1086/176550",
    journal = "Astrophys. J.",
    volume = "455",
    pages = "7--25",
    year = "1995"
}

@article{Bouhmadi-Lopez:2025spo,
    author = "Bouhmadi-Lopez, Mariam and Ibarra-Uriondo, Benat",
    title = "{Cosmological perturbations for smooth sign-switching dark energy models}",
    eprint = "2506.18992",
    archivePrefix = "arXiv",
    primaryClass = "gr-qc",
    doi = "10.1016/j.dark.2025.102129",
    journal = "Phys. Dark Univ.",
    volume = "50",
    pages = "102129",
    year = "2025"
}

@ARTICLE{1992PhR...215..203M,
       author = {{Mukhanov}, V.F. and {Feldman}, H.A. and {Brandenberger}, R.H.},
        title = "{Theory of cosmological perturbations}",
      journal = {physrep},
         year = 1992,
        month = jun,
       volume = {215},
       number = {5-6},
        pages = {203-333},
          doi = {10.1016/0370-1573(92)90044-Z}
}

@article{Akarsu:2022typ,
    author = {Akarsu, Ozgur and Kumar, Suresh and {\"O}z{\"u}lker, Emre and Vazquez, J. Alberto and Yadav, Anita},
    title = "{Relaxing cosmological tensions with a sign switching cosmological constant: Improved results with Planck, BAO, and Pantheon data}",
    eprint = "2211.05742",
    archivePrefix = "arXiv",
    primaryClass = "astro-ph.CO",
    doi = "10.1103/PhysRevD.108.023513",
    journal = "Phys. Rev. D",
    volume = "108",
    number = "2",
    pages = "023513",
    year = "2023"
}

@article{Yadav:2024duq,
    author = "Yadav, Anita and Kumar, Suresh and Kibris, Cihad and Akarsu, Ozgur",
    title = "{{\ensuremath{\Lambda}}$_{s}$CDM cosmology: alleviating major cosmological tensions by predicting standard neutrino properties}",
    eprint = "2406.18496",
    archivePrefix = "arXiv",
    primaryClass = "astro-ph.CO",
    doi = "10.1088/1475-7516/2025/01/042",
    journal = "JCAP",
    volume = "01",
    pages = "042",
    year = "2025"
}

@article{Alexandre:2023nmh,
    author = "Alexandre, Bruno and Gielen, Steffen and Magueijo, Jo{\~a}o",
    title = "{Overall signature of the metric and the cosmological constant}",
    eprint = "2306.11502",
    archivePrefix = "arXiv",
    primaryClass = "hep-th",
    doi = "10.1088/1475-7516/2024/02/036",
    journal = "JCAP",
    volume = "02",
    pages = "036",
    year = "2024"
}

@article{Anchordoqui:2023woo,
    author = "Anchordoqui, Luis A. and Antoniadis, Ignatios and Lust, Dieter",
    title = "{Anti-de Sitter {\textrightarrow} de Sitter transition driven by Casimir forces and mitigating tensions in cosmological parameters}",
    eprint = "2312.12352",
    archivePrefix = "arXiv",
    primaryClass = "hep-th",
    reportNumber = "MPP-2023-288; LMU-ASC 39/23",
    doi = "10.1016/j.physletb.2024.138775",
    journal = "Phys. Lett. B",
    volume = "855",
    pages = "138775",
    year = "2024"
}

@article{Anchordoqui:2024gfa,
    author = "Anchordoqui, Luis A. and Antoniadis, Ignatios and Lust, Dieter and Noble, Neena T. and Soriano, Jorge F.",
    title = "{From infinite to infinitesimal: Using the universe as a dataset to probe Casimir corrections to the vacuum energy from fields inhabiting the dark dimension}",
    eprint = "2404.17334",
    archivePrefix = "arXiv",
    primaryClass = "astro-ph.CO",
    reportNumber = "MPP-2024-91; LMU-ASC 05/24",
    doi = "10.1016/j.dark.2024.101715",
    journal = "Phys. Dark Univ.",
    volume = "46",
    pages = "101715",
    year = "2024"
}

@article{Anchordoqui:2024dqc,
    author = "Anchordoqui, Luis A. and Antoniadis, Ignatios and Bielli, Daniele and Chatrabhuti, Auttakit and Isono, Hiroshi",
    title = "{Thin-wall vacuum decay in the presence of a compact dimension meets the H$_{0}$ and S$_{8}$ tensions}",
    eprint = "2410.18649",
    archivePrefix = "arXiv",
    primaryClass = "hep-th",
    doi = "10.1007/JHEP07(2025)021",
    journal = "JHEP",
    volume = "07",
    pages = "021",
    year = "2025"
}

@article{Akarsu:2024nas,
    author = "Akarsu, Ozgur and Bulduk, Bilal and De Felice, Antonio and others",
    title = "{Unexplored regions in teleparallel f(T) gravity: Sign-changing dark energy density}",
    eprint = "2410.23068",
    archivePrefix = "arXiv",
    primaryClass = "gr-qc",
    reportNumber = "YITP-24-119",
    doi = "10.1103/1xd4-k91h",
    journal = "Phys. Rev. D",
    volume = "112",
    number = "8",
    pages = "083532",
    year = "2025"
}

@article{Souza:2024qwd,
    author = {Souza, Mateus S. and Barcelos, Ana M. and Nunes, Rafael C. and Akarsu, {O}zg{u}r and Kumar, Suresh},
    title = "{Mapping the LsCDM Scenario to f(T) Modified Gravity: Effects on Structure Growth Rate}",
    eprint = "2501.18031",
    archivePrefix = "arXiv",
    primaryClass = "astro-ph.CO",
    doi = "10.3390/universe11010002",
    journal = "Universe",
    volume = "11",
    number = "1",
    pages = "2",
    year = "2025"
}

@article{Akarsu:2025gwi,
    author = {Akarsu, {\"O}zg{\"u}r and Perivolaropoulos, Leandros and Tsikoundoura, Anna and Y{\"u}kselci, A. Emrah and Zhuk, Alexander},
    title = "{Dynamical dark energy with AdS-to-dS and dS-to-dS transitions: Implications for the $H_0$ tension}",
    eprint = "2502.14667",
    archivePrefix = "arXiv",
    primaryClass = "astro-ph.CO",
    month = "2",
    year = "2025",
    journal = "arXiv",
}

@article{Kumar:2025obb,
    author = "Kumar, Suresh",
    title = "{{\ensuremath{\Omega}}1{\ensuremath{\Omega}}2{\textendash}{\ensuremath{\Lambda}}CDM: A promising phenomenological extension of the standard model of cosmology}",
    eprint = "2512.19000",
    archivePrefix = "arXiv",
    primaryClass = "astro-ph.CO",
    doi = "10.1016/j.dark.2026.102248",
    journal = "Phys. Dark Univ.",
    volume = "52",
    pages = "102248",
    year = "2026"
}

@article{GelmanRubin1992,
  author  = {Gelman, Andrew and Rubin, Donald B.},
  title   = {Inference from Iterative Simulation Using Multiple Sequences},
  journal = {Statistical Science},
  volume  = {7},
  number  = {4},
  pages   = {457--472},
  year    = {1992},
  doi     = {10.1214/ss/1177011136}
}

@article{Gelman:1996ppc,
  author = {Gelman, Andrew and Meng, Xiao-Li and Stern, Hal},
  title = {Posterior predictive assessment of model fitness via realized discrepancies},
  journal = {Statistica Sinica},
  volume = {6},
  pages = {733--807},
  year = {1996}
}

@article{DESI:2025dr2bao,
    author = "Abdul-Karim, M. and others",
    collaboration = "DESI Collaboration",
    title = "{DESI DR2 Results II: Measurements of Baryon Acoustic Oscillations and Cosmological Constraints}",
    eprint = "2503.14738",
    archivePrefix = "arXiv",
    primaryClass = "astro-ph.CO",
    doi = "10.1103/tr6y-kpc6",
    journal = "Phys. Rev. D",
    volume = "112",
    number = "8",
    pages = "083515",
    year = "2025"
}

@article{DESI:2025dr2lya,
      title = {DESI DR2 results. I. Baryon acoustic oscillations from the Lyman alpha forest},
      author = "Abdul Karim, M. and others",
  collaboration = {DESI Collaboration},
  journal = {Phys. Rev. D},
  volume = {112},
  issue = {8},
  pages = {083514},
  numpages = {28},
  year = {2025},
  month = {Oct},
  publisher = {American Physical Society},
  doi = {10.1103/2wwn-xjm5},
  url = {https://link.aps.org/doi/10.1103/2wwn-xjm5},
  eprint = "2503.14739",
  archivePrefix = "arXiv",
  primaryClass = "astro-ph.CO"
}

\end{document}